\documentclass[twocolumn]{aastex63}
\usepackage[]{graphicx}

\shorttitle{size evolution in giant impacts}
\shortauthors{Matsumoto et al.}

\begin{document}

\title{Size evolution of close-in super-Earths through giant impacts and photoevaporation}

\correspondingauthor{Yuji Matsumoto}
\email{yuji.matsumoto@nao.ac.jp}

\author[0000-0002-2383-1216]{Yuji Matsumoto}
\affiliation{National Astronomical Observatory of Japan, 2-21-1, Osawa, Mitaka, 181-8588 Tokyo, Japan}
\affiliation{Institute of Astronomy and Astrophysics, Academia Sinica, Taipei 10617, Taiwan}
\author[0000-0002-5486-7828]{Eiichiro Kokubo}
\affiliation{National Astronomical Observatory of Japan, 2-21-1, Osawa, Mitaka, 181-8588 Tokyo, Japan}
\author[0000-0002-5067-4017]{Pin-Gao Gu}
\affiliation{Institute of Astronomy and Astrophysics, Academia Sinica, Taipei 10617, Taiwan}
\author[0000-0002-9958-176X]{Kenji Kurosaki}
\affiliation{Department of Physics, Nagoya University, Chikusa-ku, Nagoya 464-8602, Japan}

\begin{abstract}
The Kepler transit survey with follow-up spectroscopic observations has discovered numerous super-Earth sized planets and revealed intriguing features of their sizes, orbital periods, and their relations between adjacent planets.
For the first time, we investigate the size evolution of planets via both giant impacts and photoevaporation to compare with these observed features.
We calculate the size of a protoplanet, which is the sum of its core and envelope sizes, by analytical models.
$N$-body simulations are performed to evolve planet sizes during the giant impact phase with envelope stripping via impact shocks. 
We consider the initial radial profile of the core mass and the initial envelope mass fractions as parameters.
Inner planets can lose their whole envelopes via giant impacts, while outer planets can keep their initial envelopes since they do not experience giant impacts.
Photoevaporation is simulated to evolve planet sizes afterward.
Our results suggest that the period-radius distribution of the observed planets would be reproduced if we perform simulations in which the initial radial profile of the core mass follows a wide range of power-law distributions and the initial envelope mass fractions are $\sim0.1$.
Moreover, our model shows that the adjacent planetary pairs have similar sizes and regular spacings, with slight differences from detailed observational results such as the radius gap. 
\end{abstract}


\section{Introduction} \label{sect:intro}

Observations have reported a substantial number of low-mass and/or small-size planets, whose typical mass and size are between those of Earth and Neptune \citep[e.g.,][]{Mayor+2011, Thompson+2018}.
Hereafter, we refer to such planets as super-Earths.
Recently, the Kepler transit survey with follow-up spectroscopic observations has revealed the size and orbital period distributions of super-Earths with their orbital periods shorter than $\sim$100~days \citep[e.g.,][]{Lissauer+2011b, Petigura+2017}.
The size distribution of super-Earths in close-in orbits is bimodal with a gap of super-Earths around $2R_{\oplus}$, which is referred to as ``radius gap" \citep[e.g.,][]{Fulton+2017, Fulton&Petigura2018, Hirano+2018b, Van_Eylen+2018, Berger+2018, Berger+2020b}.
Besides, super-Earths within multiplanet systems show that adjacent planets are similar in size and their period ratios of adjacent planet pairs are similar \citep[][]{Weiss+2018, Weiss&Petigura2020}.
These features are referred to as ``peas-in-a-pod". 
It should be noted that the peas-in-a-pod patterns are pointed out to be explained by detection biases \citep{Zhu_W2020, Murchikova&Tremaine2020}.

The size of a super-Earth is significantly affected by its H/He envelope \citep[e.g.,][]{Valencia+2010,Lopez&Fortney2014}.
Close-in super-Earths or their precursors acquired H/He envelopes in situ \citep{Ikoma&Hori2012} and/or during migration in protoplanetary disks \citep[e.g.,][]{Rogers_LA+2011, Hori&Ogihara2020}.
These planets lose their envelopes while disk gas depletes \citep[e.g.,][]{Ikoma&Hori2012,Owen&Wu2016, Ginzburg+2016, Lee_EJ+2018}.
After disk gas depletion, planets experience giant impacts, which cause envelope-loss due to the shock wave \citep[e.g.,][]{Genda&Abe2003, Schlichting+2015, Liu_SF+2015, Kegerreis+2020, Denman+2020} and thermal expansion \citep{Biersteker&Schlichting2019}.
Finally, envelopes are stripped via photoevaporation \citep[e.g.,][]{Lopez&Fortney2013,Owen&Wu2013,Kurosaki+2014} and core-powered mass-loss \citep[e.g.,][]{Ginzburg+2018}, which are considered as major candidates for shaping the radius gap \citep[e.g.,][]{Owen&Wu2017,Gupta&Schlichting2019}.

The key to understanding the orbital architecture of observed super-Earths, which contains the distributions of their sizes, periods, and their ratios between adjacent planet pairs, is the H/He envelope and dynamical evolution of super-Earths.
There have been a few studies, which performed $N$-body simulations considering envelopes of planets.
\cite{Dawson+2015} and \cite{Ogihara+2018} performed $N$-body simulations taking gas accretion into account. 
Recently, \cite{Ogihara&Hori2020} and \cite{Ogihara+2020} developed a unified model, which includes gas accretion and envelope-loss via the impact shock wave and photoevaporation. 
However, in their $N$-body simulations, planetary sizes are estimated by the fixed densities of cores and envelopes. 
Moreover, their simulation samples are limited, and more systematic surveys are needed to understand the effects of the giant impact evolution on the orbital architecture of observed super-Earths.

In this paper, we examine the size evolution of planets to consider whether the orbital architecture of observed super-Earths can be reproduced.
Our simulations are the first to provide a more consistent size evolution through the giant impact stage and the subsequent long-term size evolution of planets.
We deal with the initial envelope fraction of protoplanets as a parameter, which enables us to ignore the complexities in the gas accretion and envelope-loss processes during disk gas depletion.
Our simulations proceed with the $N$-body part, followed by the photoevaporation part. 
In both parts, we track planetary sizes, which are estimated according to the analytical model based on \cite{Owen&Wu2017}.
The details of our model are described in Section \ref{sect:model}.
Our simulation results are presented in Section \ref{sect:results}.
We compare the orbital architecture of formed planets to those of observed super-Earths in Section \ref{sect:comparison}.
We present our conclusions in Section \ref{sect:conclusion}.

\section{Model} \label{sect:model}

We consider the size evolution of planets in $N$-body simulations and subsequent photoevaporation simulations.
Our parameters are the mass and envelope fraction distribution of the initial protoplanets.
We perform simulations for $10^7$~yr in the $N$-body simulation part and for $10^9$~yr for the photoevaporation simulation part.

\subsection{Initial Condition} \label{sect:model_init}

We start the simulations with protoplanets around 1 solar mass star. 
Our protoplanet model is based on the recent minimum-mass extrasolar nebula (MMEN) disk profile for the multitransiting planets \citep{Dai_F+2020}.
Each protoplanet possesses a solid core of one isolation mass \citep[][]{Kokubo&ida2000,Kokubo&Ida2002} and a H/He envelope.
The solid surface density distributions of cores are given by 
\begin{eqnarray}
	\Sigma = \Sigma_1 \left( \frac{a}{1\mbox{~au}} \right)^{-p_{\rm init}},
\end{eqnarray}
where the surface density at 1~au is $\Sigma_1=50\mbox{~g~cm}^{-2}$.
The power-law index of the radial profile $p_{\rm init}$ is our parameter, $p_{\rm init}=1, 3/2, 2$.
These values are taken from the MMEN disk profile \citep{Dai_F+2020}, where $\Sigma_1=50^{+33}_{-20} \mbox{~g~cm}^{-2}$ and $p=1.75\pm0.07$, since we aim to reproduce the orbital architecture of super-Earths observed by the transit survey.
Instead of $\Sigma_1$, we parameterize $p_{\rm init}$, which provides the variety of the mass and size of planets at $\sim0.1$~au.
The values of $p_{\rm init}$ are considered as a parameter because the surface density profile associated with planets becomes shallower as the inner protoplanets grow and scatter \citep[][]{Moriarty&Ballard2016,Matsumoto&Kokubo2017, Matsumoto+2020}. 

The protoplanets are initially located at $a=(0.05~{\rm au},~1~{\rm au})$, where $a$ is the semimajor axis of a protoplanet. 
The protoplanets are placed with orbital separations of 10 mutual Hill radii of cores, 
\begin{eqnarray}
	b_{\rm init} = 10 r_{\rm H,c}
	= 10 \left( \frac{2M_{\rm c}}{3M_{\odot}}\right)^{1/3}a,
\end{eqnarray}
where $M_{\rm c}$ is the core mass of a protoplanet and $M_{\odot}$ is the solar mass.
We ignore the contribution of the envelope mass to the orbital separation since envelopes are much less massive than cores.
The orbital separations of 10 mutual Hill radii are also taken from the MMEN disk model in \cite{Dai_F+2020}\footnote{
The orbital separation of protoplanets after resonant capture induced by Type I migration and orbital repulsion is about 10 Hill radii \citep[e.g.,][]{Ogihara&Ida2009, Ida&Lin2010}, although this depends on the migration timescale \citep[e.g.,][]{Ida&Lin2010, Ogihara&Kobayashi2013}.
}.
The mass of a protoplanet core is given by
\begin{eqnarray}
	M_{\rm c} &\simeq& 2\pi a b_{\rm init} \Sigma \nonumber \\
	&\simeq& 1.8 M_{\oplus}
	\left( \frac{b_{\rm init}}{10 r_{\rm H,c}} \right)^{3/2}
	\left( \frac{\Sigma_1}{50 \mbox{~g~cm}^{-2}} \right)^{3/2}
	\nonumber\\ &&\times
	\left( \frac{a}{1\mbox{~au}} \right)^{(3/2)(2-p_{\rm init})}
	,
	\label{eq:Miso}
\end{eqnarray}
where $M_{\oplus}$ is the Earth mass.
The initial eccentricities ($e$) and inclinations ($i$) are given by the Rayleigh distribution with the dispersions $\langle e^2\rangle^{1/2} = 2\langle i^2\rangle^{1/2} =r_{\rm H,c}/a$.

The initial fraction of the envelope mass to the core mass $X_{\rm init}=M_{\rm env}/M_{\rm c}$ is the other of our parameters, and $X_{\rm init} \lesssim 0.1$, which has little effect on the Hill radius of a protoplanet. 
We consider two models for the initial envelope mass function, which are the constant $X_{\rm init}$ model and the mass-dependent $X_{\rm init}$ model.
In the constant $X_{\rm init}$ model, each protoplanet simply has the same $X_{\rm init}$ value, which is $X_{\rm init}=0.1, 0.1^{3/2}, 0.1^{2}$.
This simple model is useful for considering the effects of giant impacts and photoevaporation on formed planets. 
In the mass-dependent $X_{\rm init}$ model, the initial envelope mass fraction is given by $X_{\rm init}=0.1(M_{\rm c}/M_{\oplus})$.
In this model, the outer massive protoplanets have larger envelope fractions in the $p_{\rm init}=1$ and 3/2 models.
When $p_{\rm init}=2$, protoplanets have the constant envelope fractions, $X_{\rm init}\simeq0.18$.
We note that this model is still simple to consider the realistic $X_{\rm init}$, which is not fully understood and is expected to depend intricately on several parameters such as the protoplanetary mass, temperature, gas density, envelope dust component \citep[e.g.,][]{Ikoma&Hori2012, Bodenheimer&Lissauer2014, Owen&Wu2016, Ginzburg+2016, Lee_EJ+2018}.
The realistic $X_{\rm init}$ will be obtained if we consider the formation of protoplanets and the evolution of their envelopes during disk gas depletion in our future study.

We define the envelope mass fraction of a planet without envelope-loss, $X_{\rm NEL}$.
Considering the conservation of the envelope mass, $X_{\rm NEL}$ is given by the total envelope mass divided by the total core mass of its component initial protoplanets, 
\begin{eqnarray}
	X_{\rm NEL} = \frac{\sum_{i} X_i M_{{\rm c},i} }{\sum_{i} M_{{\rm c},i} }, 
\end{eqnarray}
where the envelope and core mass from $i$-th initial protoplanets are given by $X_i M_{{\rm c},i}$ and $M_{{\rm c},i}$.
This fraction is simply equal to $X_{\rm init}$ in the constant $X_{\rm init}$ model.
In the mass-dependent $X_{\rm init}$ model, 
\begin{eqnarray}
	X_{\rm NEL} = \frac{\sum_{i} 0.1M_{{\rm c},i}^2/M_{\oplus} }{\sum_{i} M_{{\rm c},i} }.
\end{eqnarray}

\subsection{Planet Size} \label{sect:model_size}

We derive the planetary size based on the minimal analytical model derived by \cite{Owen&Wu2017}.
The details and the way on how we calculate in our simulations are summarized as follows.

The planetary size is given by the summation of the core size and the envelope size ($R_{\rm p}=R_{\rm c}+R_{\rm env}$).
The size of a core with an Earth-like composition is described by a power-law \citep[][]{Valencia+2006,Lopez&Fortney2014}, 
\begin{eqnarray}
	R_{\rm c} = \left( \frac{M_{\rm c}}{M_{\oplus}} \right)^{1/4} R_{\oplus},
	\label{eq:Rc}
\end{eqnarray}
where $R_{\oplus}$ is the Earth radius.

The envelope of a planet can be divided into two parts, the outer radiative atmosphere and the inner convective region.
The radiative atmosphere is typically much thinner than the convective region \citep{Owen&Wu2017}.
We neglect the size of the radiative atmosphere, and the envelope size is given by the size of the convective region.
Considering the hydrostatic equilibrium, the density profile in the convective region is 
\begin{eqnarray}
	\rho_{\rm env} &=& 
	\rho_{\rm rcb} \left[ 1+ \nabla_{\rm ab} \left( \frac{{\rm G}M_{\rm c}}{c_{\rm s}^2R_{\rm p}}\right) \left( \frac{R_{\rm p}}{R}-1 \right) \right]^{1/(\gamma-1)}
	\nonumber\\
	&\simeq & 
	\rho_{\rm rcb} \left[ \nabla_{\rm ab} \left( \frac{{\rm G}M_{\rm c}}{c_{\rm s}^2R_{\rm p}}\right) \left( \frac{R_{\rm p}}{R}-1 \right) \right]^{1/(\gamma-1)},
\end{eqnarray}
where $\rho_{\rm rcb}$ is the density at the radiative-convective boundary, $\nabla_{\rm ab}=(\gamma-1)/\gamma$ and $\gamma=5/3$ are the adiabatic gradient and index, ${\rm G}$ is the gravitational constant, and $c_{\rm s}$ is the sound speed at the radiative-convective boundary.
The sound speed at the radiative-convective boundary is roughly given by $c_{\rm s}=\sqrt{k_{\rm B} T_{\rm eq}/m_{\rm g} }$, where $m_{\rm g}$ is the mean molecular mass ($m_{\rm g}=3.9\times10^{-24}$~g), $k_{\rm B}$ is the Boltzmann constant, and $T_{\rm eq}$ is the equilibrium temperature, which roughly expresses the temperature at the radiative-convective boundary.
The density at the radiative-convective boundary is obtained under the equilibrium of the temperature gradient between the radiative and convective regions, 
\begin{eqnarray}
	\rho_{\rm rcb}
	&\approx& 
	\left( \frac{ m_{\rm g} }{ k_{\rm B} } \right)
	\left(
		\nabla_{\rm ab} \frac{I_2}{I_1}
		\frac{64\pi\sigma T_{\rm eq}^{3} R_{\rm p} \tau_{\rm KH} }{3\kappa_0 M_{\rm env}}
	\right)^{1/(\alpha+1)},
\end{eqnarray}
where $\sigma$ is the Stefan--Boltzmann constant. 
The opacity is given by $\kappa=\kappa_0(\rho_{\rm rcb}/10^{-3} \mbox{~g~cm}^{-3})^{\alpha}$, where $\kappa_0=0.1\mbox{~cm}^{2}\mbox{~g}^{-1}$ and $\alpha=0.6$ \citep{Freedman+2008,Gupta&Schlichting2019}.
The Kelvin--Helmholtz timescale $\tau_{\rm KH}$ is $\tau_{\rm KH} = \max{(10^8\mbox{~yr}, t)}$.
Since the Kelvin--Helmholtz timescale is related to the cooling timescale, $\rho_{\rm rcb}$ begins to increase and envelopes begin to shrink when $t>10^8\mbox{~yr}$.
We note that this Kelvin--Helmholtz timescale does not take into account any heat deposit due to giant impacts, whose dependencies on the impact velocity and angle are still not clear.
The dimensionless integrals, $I_1$ and $I_2$, are
\begin{eqnarray}
	I_1(R_{\rm c}/R_{\rm p},\gamma) &=& \int_{R_{\rm c}/R_{\rm p}}^{1}
	x
	\left( x^{-1}-1 \right)^{1/(\gamma-1)} dx, 
	\\
	I_2(R_{\rm c}/R_{\rm p},\gamma) &=& \int_{R_{\rm c}/R_{\rm p}}^{1}
	x^2 
	\left( x^{-1}-1 \right)^{1/(\gamma-1)} dx,
\end{eqnarray}
which depend on $R_{\rm p}$.

The envelope mass is expressed by 
\begin{eqnarray}
	M_{\rm env} &=& 
	\int_{R_{\rm c}}^{R_{\rm p}} 4\pi R^2 \rho_{\rm env} dR \nonumber \\
	&\simeq& 
	4\pi \rho_{\rm rcb} R_{\rm p}^{3}
	\left( \nabla_{\rm ab} \frac{{\rm G}M_{\rm c}}{c_{\rm s}^2R_{\rm p}}\right)^{1/(\gamma-1)} I_2.
\end{eqnarray}
In our simulations, we iteratively solve this integral and derive $R_{\rm p}$.
The values of $I_2/I_1$ and $I_2$ are taken from the tables, which we make at first, and interpolated.
We derive $R_{\rm p}$ at the first step and each collision in $N$-body simulations and at every step in photoevaporation simulations.

\subsection{Impact Erosion} \label{sect:model_col}

In $N$-body simulations, we consider envelope-loss via giant impact shock waves.
We adopt an empirical envelope-loss rate obtained from smoothed particle hydrodynamics simulations \citep[][]{Kegerreis+2020}.
Given a collision between $i$-th and $j$-th protoplanets with the modified specific impact energy $Q$,
\begin{eqnarray}
	Q &=& \frac{1}{2} (1-b)^2 (1+2b) \frac{\mu}{ M_i + M_j } v_{\rm col}^2,
\end{eqnarray}
where $b$ is the impact parameter, $\mu$ is the reduced mass, and $v_{\rm col}$ is the collision velocity, the envelope-loss rate is 
\begin{eqnarray}
	X_{\rm col} &\approx& 
	7.72\times 10^{-6} \left( Q/\mbox{J~kg}^{-1} \right)^{0.67}
	.
\end{eqnarray}
The impact envelope-loss rate is high when a collision is head-on ($b=0$) and a collision velocity is high.

\subsection{Photoevaporation} \label{sect:model_PE}

The envelope-loss rate via photoevaporation under the energy-limited approximation is given by
\begin{eqnarray}
	{\dot M}_{\rm XUV} &=& \epsilon_{\rm PE} \frac{\pi R_{\rm p}^3 L_{\rm XUV}}{ 4\pi a^2 {\rm G}M_{\rm p} K_{\rm tide} }, 
\end{eqnarray}
where $\epsilon_{\rm PE}=0.1$ is the heating efficiency \citep[][]{Lopez&Fortney2013}, $K_{\rm tide}$ is the correction factor due to the stellar tide \citep[][]{Erkaev+2007}, and the XUV luminosity from the star is 
\begin{eqnarray}
	L_{\rm XUV} 
	&=& \left\{
		\begin{array}{ll}
			L_{\rm sat} & (t< t_{\rm sat} )\\
			L_{\rm sat} \left( \frac{t}{t_{\rm sat}} \right)^{-1-a_0} & (t\geq t_{\rm sat} ),
		\end{array}
		\right.
\end{eqnarray}
where $L_{\rm sat}=10^{-3.5}L_{\odot}$, $a_0=0.5$, and $t_{\rm sat}=100~\mbox{Myr}$ \citep[][and references therein]{Owen&Wu2017}.

\subsection{Integration} \label{sect:model_integral}

\begin{figure}
	\plotone{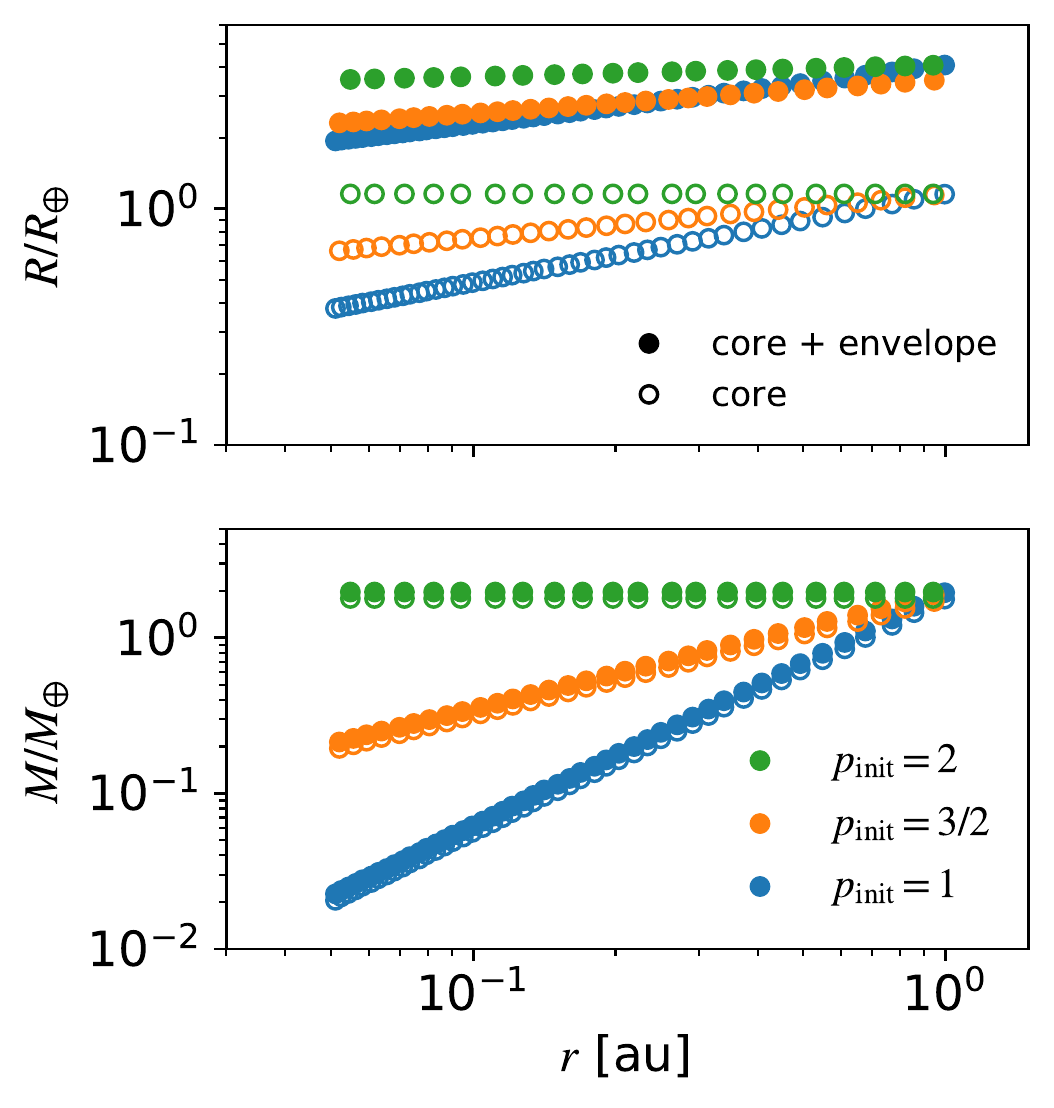}
	\caption{
		The initial size and mass distributions of protoplanets when $X_{\rm init}=0.1$.
		}
		\label{fig:rMR_init}
\end{figure}

First of all, we perform $N$-body simulations using the fourth-order Hermite scheme \citep[][]{Makino&Aarseth1992,Kokubo&Makino2004} with the hierarchical timestep \citep[][]{Makino1991}.
The initial size and mass distributions in the $X_{\rm init}=0.1$ models are shown in Figure \ref{fig:rMR_init}.
In each collision, we generate the merged body in the following manner: the core mass of the merged body is assumed to be the summation of those of the merging bodies; the envelope mass of the merged body is derived by the total envelope mass and impact erosion (Section \ref{sect:model_col}); the size of the merged body is calculated (Section \ref{sect:model_size}).
These simulations last until $10^7$~yr.
This timescale is shorter than $\tau_{\rm KH}$, and this is why we do not need to calculate a planetary size at each step.

Subsequently, we perform photoevaporation simulations. 
In these simulations, we focus on the size evolution of planets. 
We calculate envelope-loss via photoevaporation (Section \ref{sect:model_PE}) and planetary sizes with fixed orbital elements.
These simulations last until $10^9$~yr.
In each parameter set of ($X_{\rm init}$, $p_{\rm init}$), we perform 20 simulations of $N$-body and photoevaporation.

\section{Results} \label{sect:results}

\subsection{N-body simulations}\label{sect:result_N}

\subsubsection{Example run}\label{sect:N_typical}

\begin{figure}
	\includegraphics[width=\linewidth]{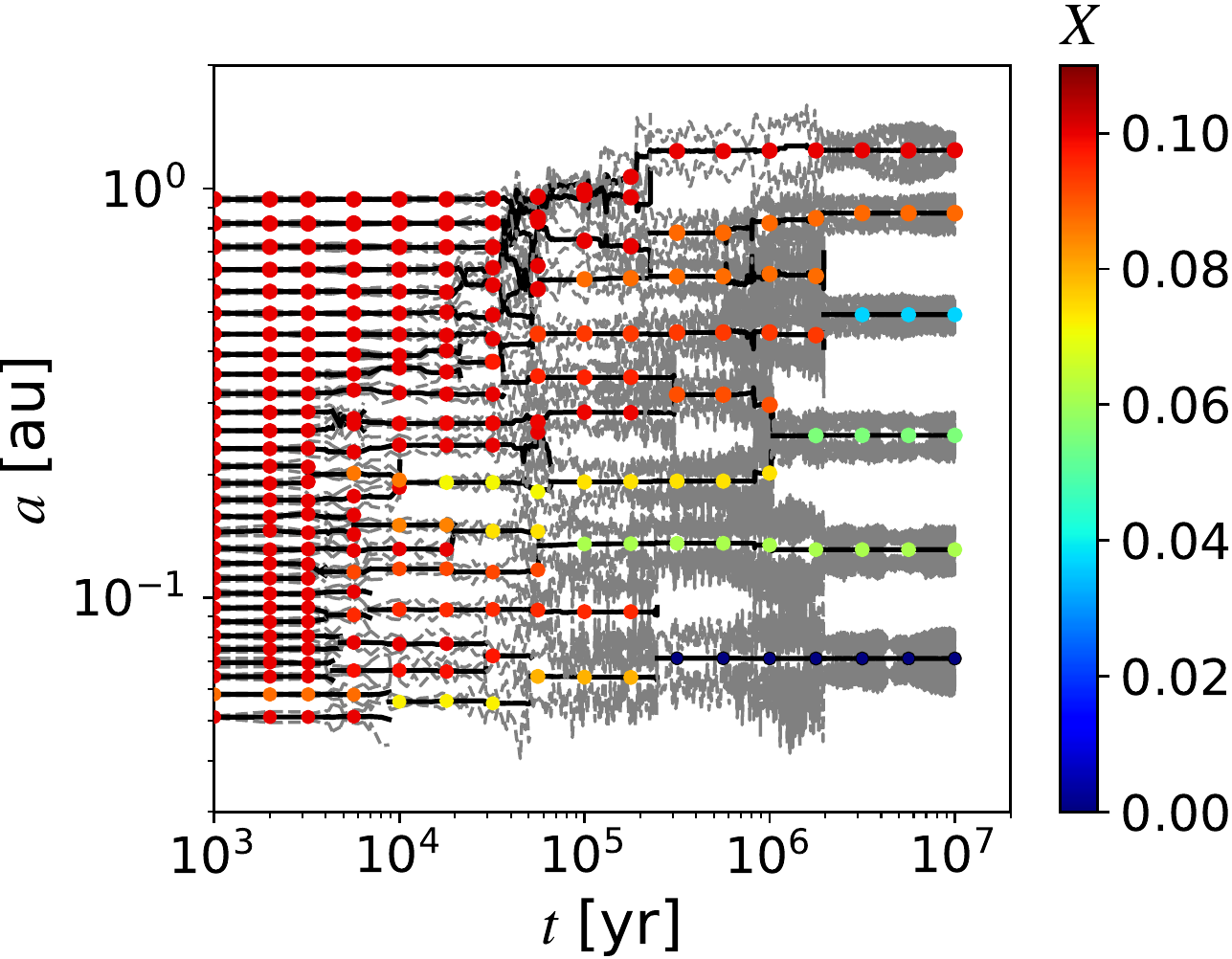}
	\caption{
		Orbital evolution of protoplanets in an example run of the $(X_{\rm init}$, $p_{\rm init})=(0.1, 3/2)$ model is shown.
		Their semimajor axes are plotted as solid black lines, and their pericenter and apocenter distances are dashed gray lines.
		The colors of the circles represent their envelope fractions ($X$).
		}
		\label{fig:t_vs_aRX}
\end{figure}

We show the orbital evolution of protoplanets in the case that the initial envelope fraction is fixed, $X_{\rm init}=0.1$, to verify the effect of impact erosions on the envelope fraction.
Figure \ref{fig:t_vs_aRX} shows the orbital evolution in an example run of the $(X_{\rm init}$, $p_{\rm init})=(0.1, 3/2)$ model. 
In this figure, we also plot the envelope fractions of protoplanets.
Collisions occur at inner orbits, $\lesssim 0.1$~au, at first. 
As collisions occur, protoplanets lose their envelopes.
In the first $10^5$~yr, 20 collisions occur. 
The envelope-loss fractions via impact erosions are $X_{\rm col}<0.24$, and their average value is $\langle X_{\rm col} \rangle =0.093\pm 0.075$.
Protoplanets typically lose about 10\% of their envelopes.
These low envelope-loss fractions are due to grazing and low-velocity collisions.
These collisions, which are grazing and whose velocities are around their escape velocities ($v_{\rm esc}$), are consistent with previous studies in which planetary envelopes are not included \citep[e.g.,][]{Kokubo&Ida2007, Raymond+2009, Stewart&Leinhardt2012}.
In the first $10^5$~yr, escape velocities are low, since masses of protoplanets are small ($\lesssim M_{\oplus}$) and sizes are large ($\gtrsim2R_{\oplus}$), and the envelope-loss fractions via collisions are, therefore, low ($X_{\rm col}\simeq 0.1$).

\begin{figure}
	\includegraphics[width=\linewidth]{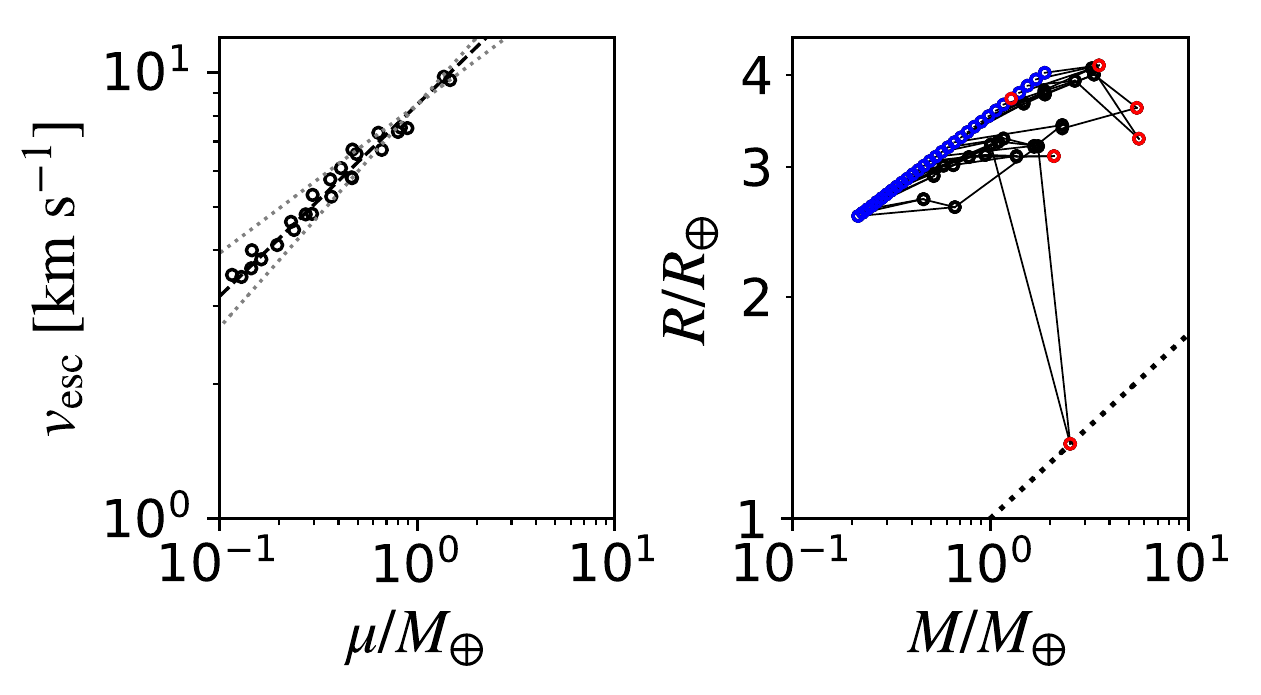}
	\caption{
		Escape velocities of protoplanets as a fucntion of the reduced mass ($\mu$) in collisions (left) and their growth on the $M$-$R$ plane (right).
		Left: 
		The dashed line is the fitting line, $v_{\rm esc}\propto M^{0.43}$, and the dotted lines are $v_{\rm esc}\propto M^{1/2}$ and $v_{\rm esc}\propto M^{1/3}$ lines, where the $v_{\rm esc}$ values are the same as that of the fitting line at $\mu=M_{\oplus}$.
		Right: 
		The initial protoplanets are shown in the blue circles and the formed planets are in the red circles.
		The merged bodies are connected with their precursors by the solid line.
		The dotted line is the mass-radius relationship of the core.
		}
		\label{fig:MRv_esc}
\end{figure}

The envelope-loss fractions via impact erosions become higher when $t>10^5$~yr, $\langle X_{\rm col} \rangle =0.42\pm 0.38$.
In particular, whole envelopes are lost in the collision at $2.5\times10^5$~yr between the innermost and second innermost protoplanets.
Higher erosion rates arise due to high collision velocities \citep[][Section \ref{sect:model_col}]{Kegerreis+2020}.
Collision velocities increase as protoplanets grow due to higher escape velocities.
Our results show that the power-law dependence of the escape velocity on the reduced mass is 0.43 in the $X_{\rm init}=0.1$ models (the left panel of Figure \ref{fig:MRv_esc}). 
The dependence of the escape velocity on the reduced mass reflects the size dependence of protoplanets on their masses.
The power-law indices are 1/3 when the density of a protoplanet is fixed and 1/2 when the planetary size is independent of the planetary mass.
The size dependence on the protoplanet mass is weak since the size of a protoplanet is given by its envelope and is weakly affected by the mass growth (the right panel of Figure \ref{fig:MRv_esc}).
Also, collision velocities increase as eccentricities and inclinations increase due to scattering and collisions \citep[][]{Matsumoto+2015, Matsumoto&Kokubo2017}.
Accordingly, the collision velocity increases as protoplanets grow.

\begin{figure}
	\plotone{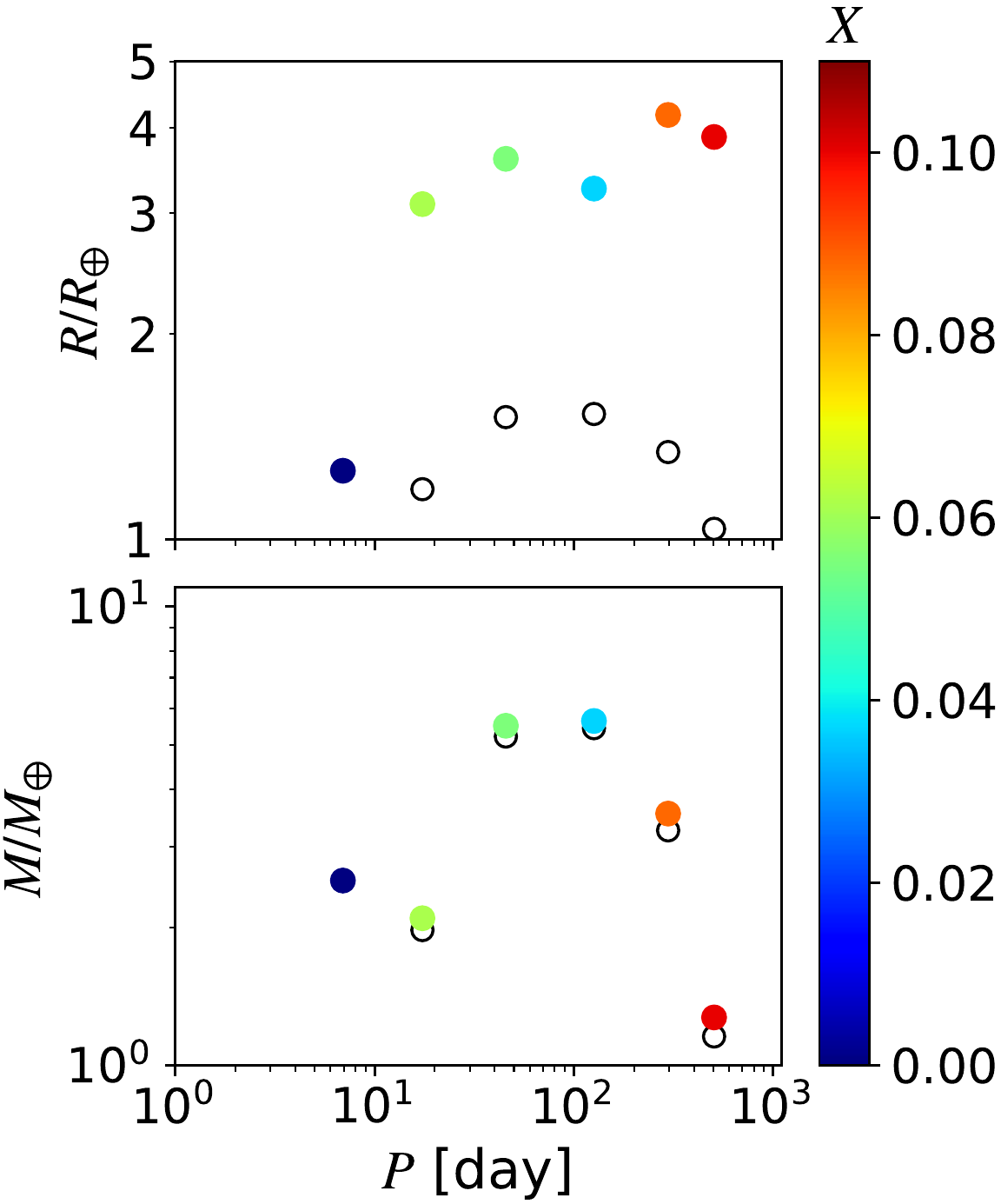}
	\caption{
		Planets at $10^7$~yr on the $P$--$R$ (top) and $P$--$M$ (bottom) planes.
		The filled circles are those of cores and envelopes, and the open circles are those of cores.
		The colors of the filled circles are their envelope fractions.
		}
		\label{fig:final_PMR}
\end{figure}

Formed planets are shown in Figure \ref{fig:final_PMR}.
Their masses are $2.5M_{\oplus}$, $2.1M_{\oplus}$, $5.5M_{\oplus}$, $5.6M_{\oplus}$, $3.5M_{\oplus}$, and $1.2M_{\oplus}$ from the innermost planet, respectively.
The formed planetary cores are similar in size since the core sizes are almost given by $M^{1/4}$.
In contrast, the sizes of planets are not similar.
The innermost planet is $1.3R_{\oplus}$, while the other planets are larger than $3R_{\oplus}$.
The size of the innermost planets is the smallest since it does not have an envelope.
The outermost planet has the second largest size although its mass is the smallest.
This planet keeps the initial envelope since it does not experience any collisions.
The final envelope fractions are 0, 0.061, 0.055, 0.037, 0.088, and 0.1.

\subsubsection{Example model}\label{sect:N_std}

\begin{figure*}
	\plotone{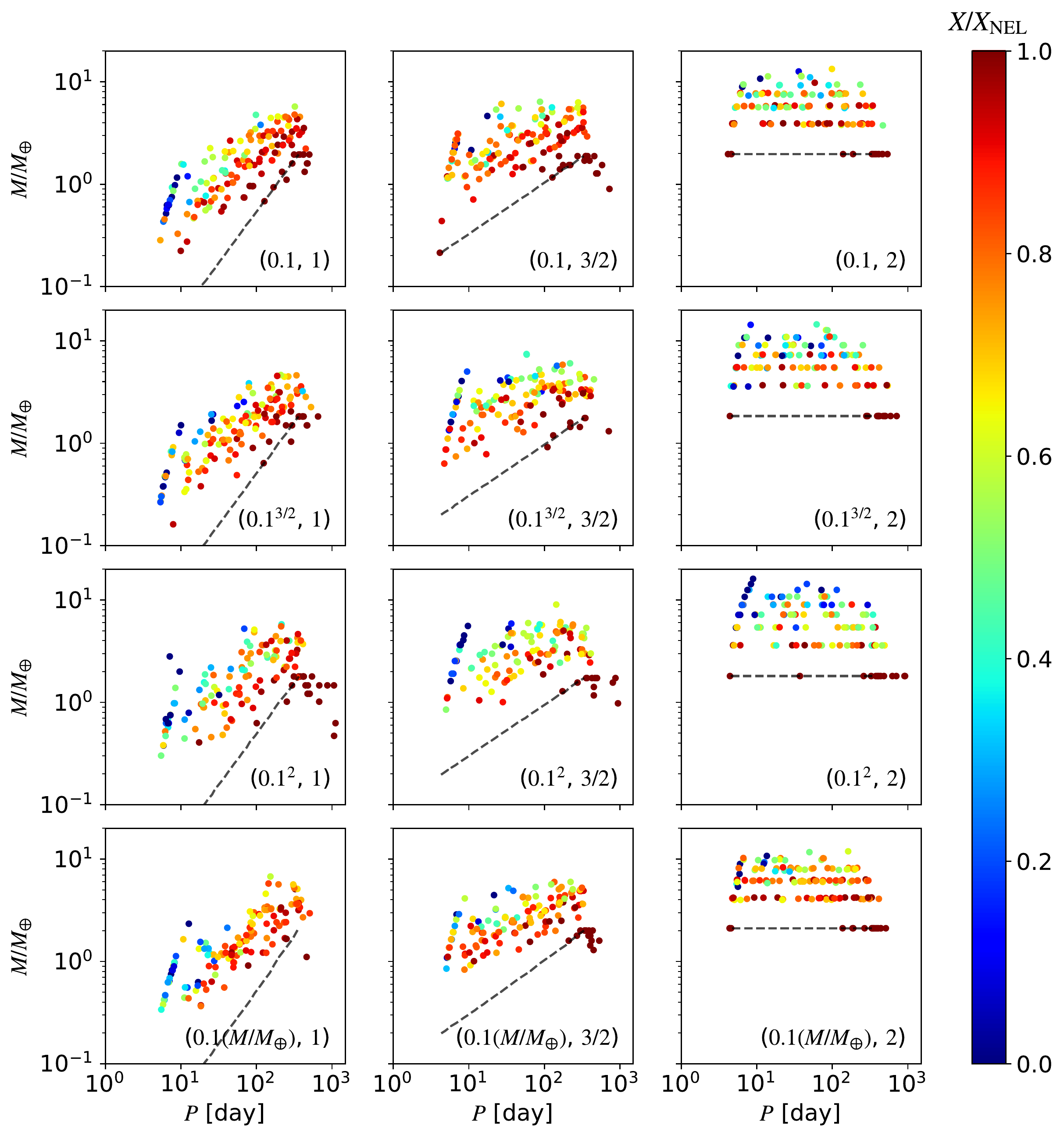}
	\caption{
		The planetary mass and period distributions at $10^7$~yr.
		We plot planets in each ($X_{\rm init}$, $p_{\rm init}$) model.
		The colors of the symbols represent the values of the envelope fraction normalized by $X_{\rm NEL}$. 
		The dashed lines are the initial distributions of protoplanets.
		}
		\label{fig:final_PMX_XM}
\end{figure*}

\begin{figure*}
	\plotone{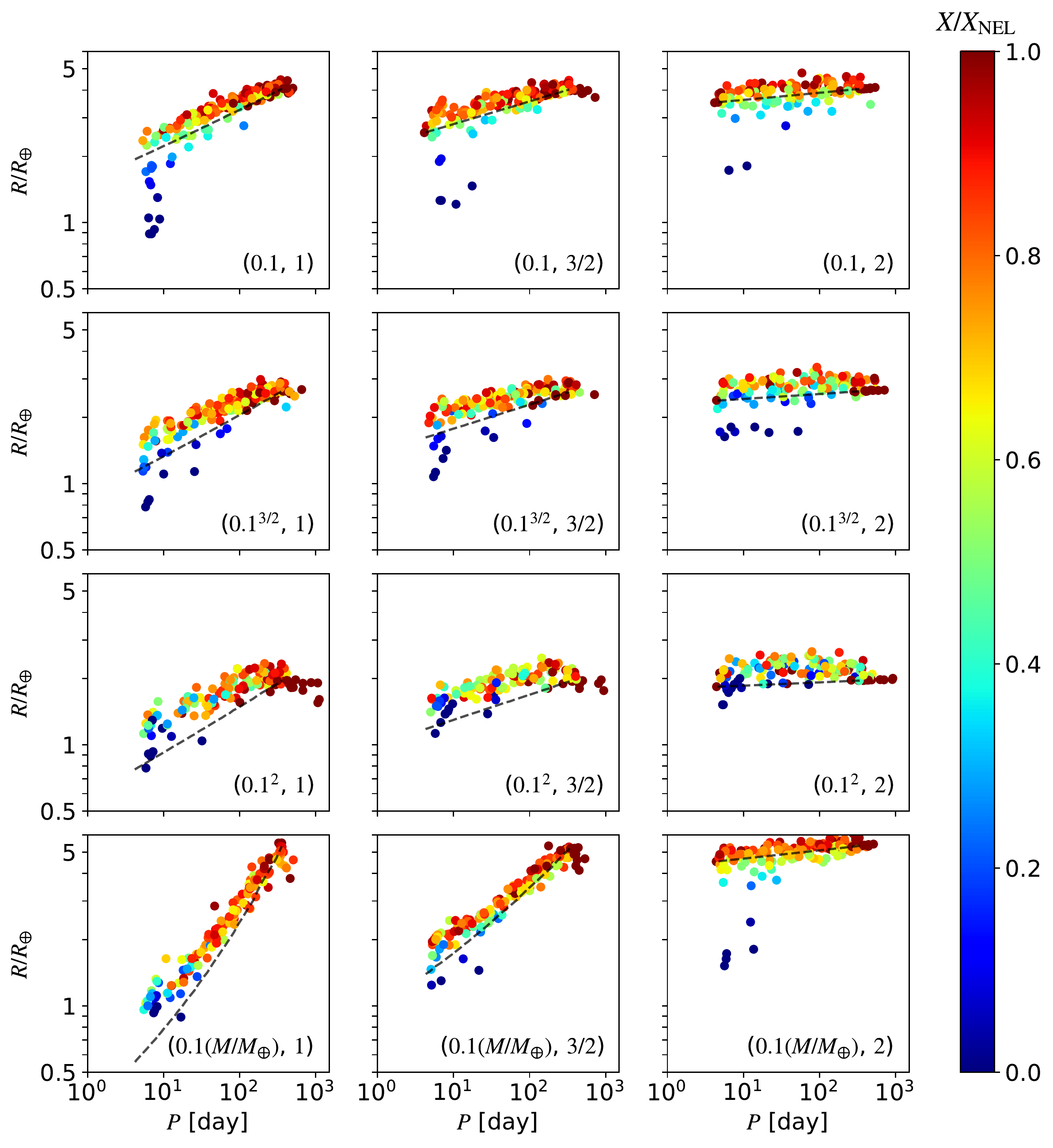}
	\caption{
		The same as Figure \ref{fig:final_PMX_XM}, but for the planetary size and period distributions at $10^7$~yr (after $N$-body simulations).
		}
		\label{fig:final_PRX_XM}
\end{figure*}

\begin{figure*}
	\plotone{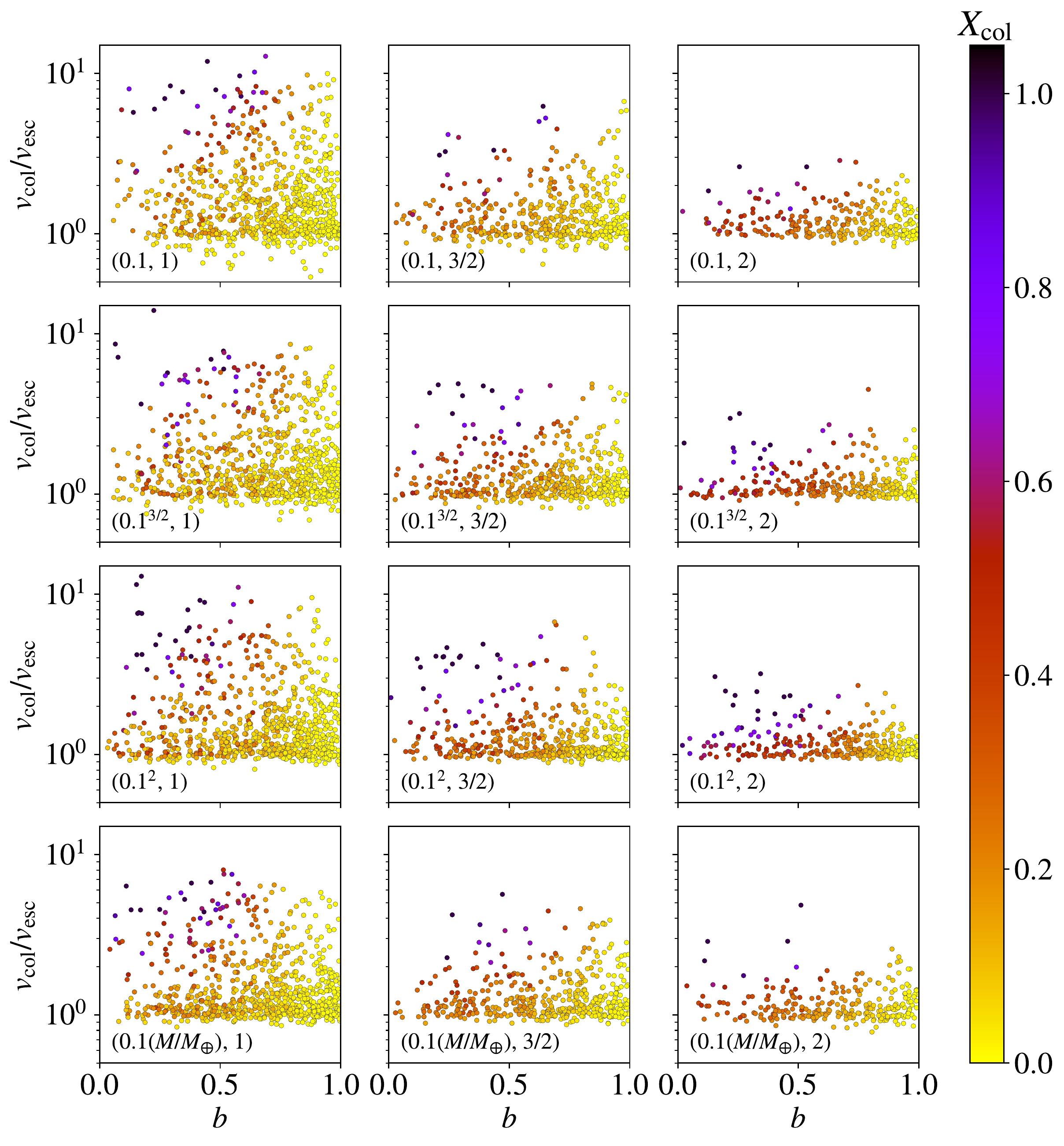}
	\caption{
		The collision velocity normalized by the escape velocity ($v_{\rm col}/v_{\rm esc}$) is plotted against the impact parameter ($b$).
		Point colors represent the envelope-loss rate ($X_{\rm col}$).
		}
		\label{fig:b_vcol_XM_mat}
\end{figure*}

The mass and size distributions of the formed planets are shown in Figures \ref{fig:final_PMX_XM} and \ref{fig:final_PRX_XM}.
Also, the collision velocities and impact parameters are shown in Figure \ref{fig:b_vcol_XM_mat}.
First, we focus on the $(X_{\rm init},p)=(0.1, 3/2)$ model.
In 20 runs of this model, 156 planets are formed.
The mass distribution shows that these planets are composed of the inner less-massive planets, the middle massive planets, and the outer less-massive planets.
Bare-core planets, which have smaller radii than other planets, and initial-envelope planets, which are less massive than other planets, are found in these distributions.
There are four bare planets, which have 2.2 -- 4.6$M_{\oplus}$ and 1.2 -- 1.5$R_{\oplus}$, and they are located at inner orbits (6.6 -- 18~day orbital periods).
These planets lose their whole envelopes in the collisions at $t>10^5$~yr due to the high-velocity collisions (Section \ref{sect:N_typical}, Figure \ref{fig:b_vcol_XM_mat}).
While high-velocity collisions, which are higher than $2 v_{\rm esc}$, are only about 13\% of all collisions, this small fraction of collisions contributes to forming bare planets.

Initial envelope planets are more common than bare planets, and 19 initial-envelope planets are formed.
Except for one planet, which is located at 0.05~au (4.1~day period), they are located at 0.4 -- 1.6~au (94 -- 726~day periods).
These planets keep their initial masses and sizes.

The other 133 planets typically experience $3.2\pm2.2$ collisions and have 76\%~$\pm$~18\% of their initial envelopes.
Most collisions in the $(X_{\rm init},p)=(0.1, 3/2)$ model are low velocity ($v_{\rm col}/v_{\rm esc}<2$) and oblique ($b>b_{\rm crit}\simeq0.5$, where $b_{\rm crit}$ is the critical impact parameter given by the size ratio between the target and the total \citep[][]{Asphaug2010,Leinhardt&Stewart_ST2012}). 
Due to large envelopes, these planets have 2.5 -- 4.4$R_{\oplus}$.
Their size distribution is similar to the initial size distribution since their sizes are given mostly by their envelopes.
Owing to the core growth, the final sizes are slightly larger than the initial sizes.

\subsubsection{Parameter dependence}\label{sect:N_par}

\begin{figure*}
	\plotone{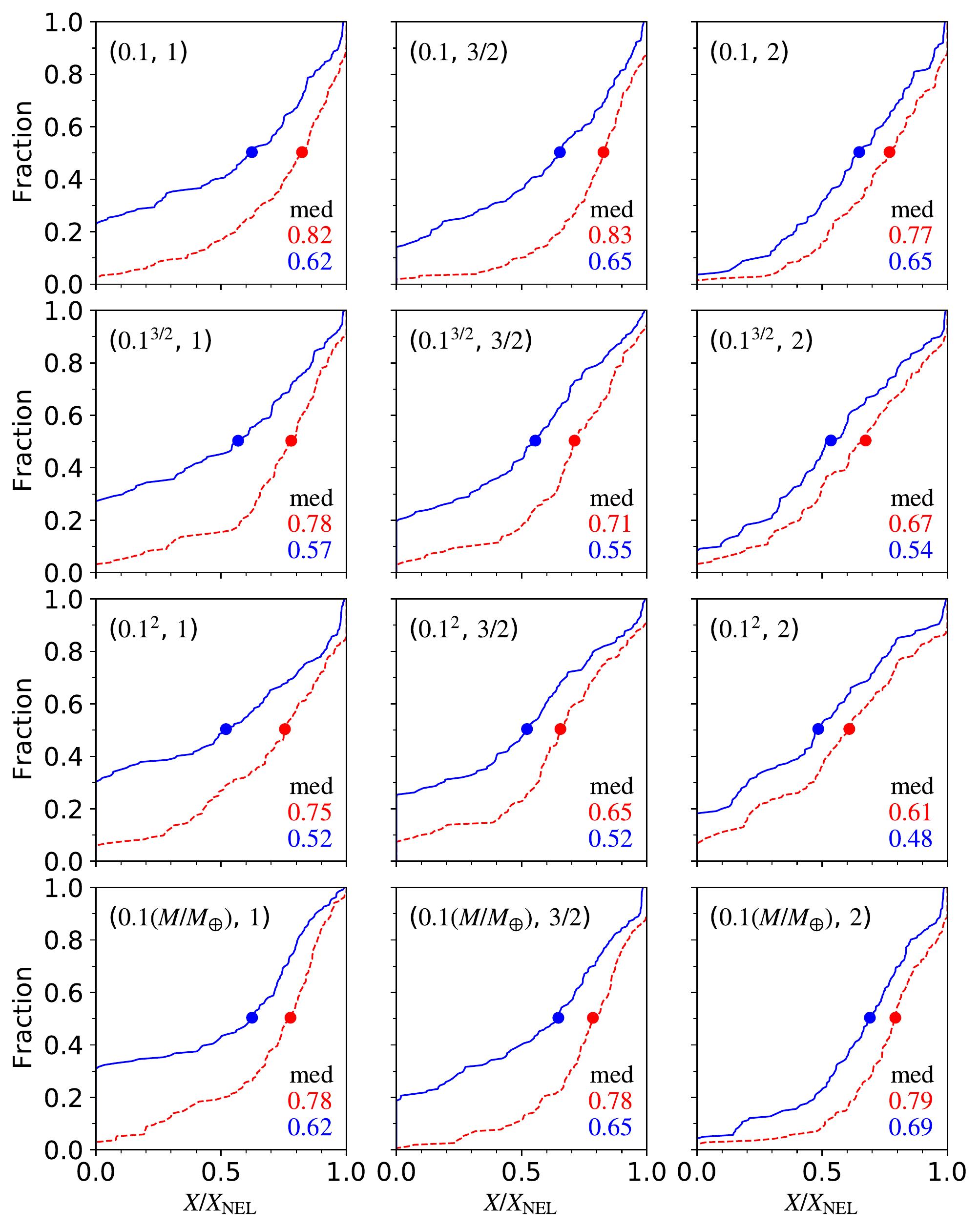}
	\caption{
		The cumulative fractions of $X/X_{\rm NEL}$ in each ($X_{\rm init}$, $p_{\rm init}$) model.
		The red dashed lines are those values at $10^7$~yr, and the blue solid lines are those at $10^9$~yr.
		The circles indicate the medians, which are denoted at the bottom right of each panel.
		}
		\label{fig:X_Xinit_cum_GIPE}
\end{figure*}

First, we explain the results in the constant $X_{\rm init}$ models.
The mass distributions of formed planets are basically similar between the same $p_{\rm init}$ models.
Although the giant impact growth makes the mass distribution shallower and the final mass distributions are different from the initial ones due to the quick growth of inner protoplanets, the initial power-law index ($p_{\rm init}$) affects the mass distribution of the formed planets (Figure \ref{fig:final_PMX_XM}). 
As $p_{\rm init}$ increases, more massive planets are formed since protoplanets grow via collisions with adjacent protoplanets at close-in orbits.
The mass growth of protoplanets is not affected by their envelope fractions at inner orbits since they tend to collide immediately after the orbital crossing even without envelopes \citep[][]{Matsumoto&Kokubo2017}.

In contrast, their size distributions are different in each model and are similar to their initial distributions (Figure \ref{fig:final_PRX_XM}).
As opposed to the mass distributions, the size distributions are not relaxed through giant impacts since the core size dependence on the core mass is weak (Equation (\ref{eq:Rc})).
The sizes of most of the formed planets are determined by their envelope fractions.
Figure \ref{fig:X_Xinit_cum_GIPE} shows the cumulative distributions of the ratio of the envelope fractions to the envelope mass fraction of planets without envelope-loss ($X/X_{\rm NEL}$).
The formed planets tend to have more than $0.5X_{\rm NEL}$ envelopes, and the medians of $X/X_{\rm NEL}$ are between 0.61 and 0.83 at $10^7$~yr.
Larger core and thick envelopes contribute to large planetary sizes, which are slightly larger than the initial sizes.

Bare planets and initial-envelope planets are also found in the other $(X_{\rm init},p)$ models. 
Bare planets are located at $\sim0.1$~au ($\sim10$~day period orbits). 
Inner planets experience more collisions and scattering, which pump up their eccentricities and cause high-velocity collisions, and as a result, inner planets become bare.
This is also the reason why higher $v_{\rm col}/v_{\rm esc}$ collisions occur in smaller $p_{\rm init}$ models (Figure \ref{fig:b_vcol_XM_mat}).
In smaller $p_{\rm init}$ models, inner protoplanets are less massive and their numbers are large (Figure \ref{fig:rMR_init}), which induces more collisions and higher $v_{\rm col}/v_{\rm esc}$ collisions.
It is worth noting that more than 70\% of collisions are $v_{\rm col}/v_{\rm esc}<2$ even in the $p_{\rm init}=1$ models.
Although more collisions of high $v_{\rm col}/v_{\rm esc}$ occur in smaller $p_{\rm init}$ models, the fractions of the bare planets do not show a systematic trend with $p_{\rm init}$.
This is because the envelope-loss fractions via impact erosions depend on $v_{\rm col}$.
In the large $p_{\rm init}$ model, massive planets are formed, and their high $v_{\rm esc}$ values contribute to the envelope-loss fractions via impact erosions.
In contrast, the fractions of bare planets increase as the initial envelope fractions decrease.
The fractions of bare planets are similar in the same $X_{\rm init}$ models; i.e., 
$0.025\pm0.0025$ in the $X_{\rm init}=0.1$ models, 
$0.036\pm0.0028$ in the $X_{\rm init}=0.1^{3/2}$ models, 
$0.073\pm0.0075$ in the $X_{\rm init}=0.1^{2}$ models, respectively (Figure \ref{fig:X_Xinit_cum_GIPE}).
The reason for this tendency would be that some small envelope protoplanets cause high-velocity collisions at small impact parameters since they experience more scattering between collisions due to small cross-sections.

The medians of $X/X_{\rm NEL}$ at $10^7$~yr (red circles in Figure \ref{fig:X_Xinit_cum_GIPE}) increase as $X_{\rm init}$ increases. 
This is consistent with the fractions of bare planets. 
The medians decrease as $p_{\rm init}$ increases in the $X_{\rm init}=0.1^{3/2}$ and $0.1^2$ models, but this tendency is not clear in the $X_{\rm init}=0.1$ and mass-dependent models.
More simulations are needed to consider whether the giant impact envelope erosion is affected by the initial mass distribution.

Initial-envelope planets account for $\sim$10\% (5.8\% -- 15\%) of formed planets, and this fraction does not have systematic dependence on $p_{\rm init}$ and $X_{\rm init}$ (Figure \ref{fig:X_Xinit_cum_GIPE}).
This fraction suggests that there is a planet that does not experience collisions per one or two systems since final systems host 6 -- 8 planets in our results.
Except for the inner ones, the initial-envelope planets are located at $\geq0.4$~au ($\geq100$~day period orbits), and their medians of semimajor axes are 1.0~au, which is around the initial locations of the outermost protoplanets.

The mass-dependent $X_{\rm init}$ models show similar features.
In the $p_{\rm init}=1$ and $3/2$ models, in which inner protoplanets have small envelope fractions than outer ones, outer planets have similar sizes to the initial protoplanets since they have thick envelopes, which determine their size.
Inner planets are larger than the initial protoplanets due to the core growth.

Our results show that although the mass-period distributions of planets are similar for the same $X_{\rm init}$, $X_{\rm init}$ affects the final $X/X_{\rm NEL}$ values (Figure \ref{fig:X_Xinit_cum_GIPE}) and the fractions of the bare planets.
We suggest that previous studies, in which the size or mass evolution of protoplanets is not included \citep[e.g.,][]{Quintana+2016, Inamdar&Schlichting2016}, did not obtain a precise estimation on the envelope fraction after the giant impact stage: there are 7\% -- 18\% differences of the final $X/X_{\rm NEL}$ median values between the $X_{\rm init}=0.1$ cases and the $X_{\rm init}=0.1^2$ cases; there are 4.8\% differences of the fractions of the bare planets between the $X_{\rm init}=0.1$ cases and the $X_{\rm init}=0.1^2$ cases. 

\subsection{photoevaporation simulations}\label{sect:result_PE}

\begin{figure*}
	\plotone{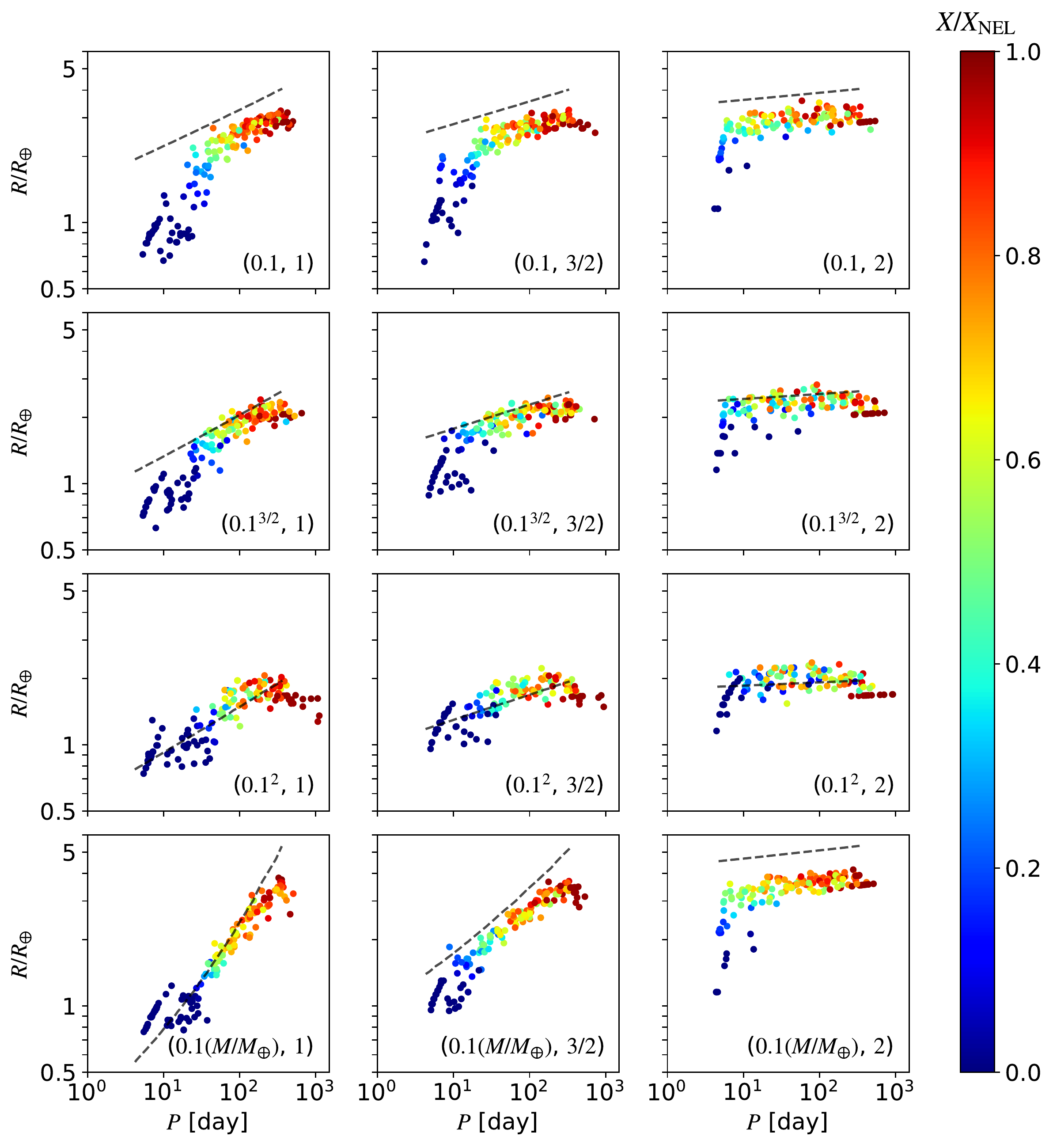}
	\caption{
		The same as Figure \ref{fig:final_PRX_XM}, but at $10^9$~yr (after photoevaporation simulations).
		}
		\label{fig:PE_PRX_XM}
\end{figure*}

Through photoevaporation and cooling contraction, the sizes of the formed planets decrease.
Figure \ref{fig:PE_PRX_XM} shows the size distributions of the formed planets at $10^9$~yr.
All planets have smaller sizes than those at $10^7$~yr.
These size reductions are more apparent on inner planets due to envelope-loss via photoevaporation.
Since photoevaporation strongly affects less massive planets, planets in the ($X_{\rm init}$, $p_{\rm init}$)=($0.1$, $1$) and ($0.1$, $3/2$) models, in particular, clearly show the distributions of the inner small envelope planets and outer large envelope planets.
More bare planets are formed in smaller $p_{\rm init}$ and smaller $X_{\rm init}$ models (Figure \ref{fig:X_Xinit_cum_GIPE}).
In the ($X_{\rm init}$, $p_{\rm init}$)=($0.1^2$, $1$) model, planets easily lose their whole envelopes, and the fraction of the bare planets is 31\% of the formed planets, which is 24\% larger than the fraction at $10^7$~yr.
In comparison, in the ($X_{\rm init}$, $p_{\rm init}$)=($0.1$, $2$) model, planets tend to keep their envelopes, the fraction of the bare planets is 4.4\% of the formed planets, which is 2.2\% larger than the fraction at $10^7$~yr.

While photoevaporation leads to more bare planets at inner orbits ($\lesssim 10$~days), outer initial-envelope planets with more than 100~day periods keep almost all their envelopes even after photoevaporation (Figures \ref{fig:X_Xinit_cum_GIPE} and \ref{fig:PE_PRX_XM}).
This indicates that observed planets that are located at $\gtrsim1$~au and outermost in a system have the almost same envelopes from the stage of disk gas depletion.

The $X/X_{\rm NEL}$ cumulative fractions after photoevaporation (blue lines in Figure \ref{fig:X_Xinit_cum_GIPE}) tends to reflect the distributions of planets on the mass and semimajor axis plane.
This tendency is strong for the $p_{\rm init}=1$ models, where inner planets are less massive, and weak for the $p_{\rm init}=2$ models, where inner planets are massive and their envelopes are not efficiently stripped by photoevaporation.
This makes the shapes of the $X/X_{\rm NEL}$ cumulative fractions in the same $p_{\rm init}$ models similar. 
They arise mainly from the mass distributions of planets in the $p_{\rm init}=1$ and $3/2$ models and result more directly from giant impacts in the $p_{\rm init}=2$ models.

We assess the contributions of giant impacts and photoevaporation to typical envelope-loss fractions focusing on the medians of $X/X_{\rm NEL}$.
The differences between the $X/X_{\rm NEL}$ median values at $10^9$~yr and those at $10^7$~yr ($\Delta X/X_{\rm NEL}$) tend to be large in small $p_{\rm init}$ models: 
$0.15<\Delta X/X_{\rm NEL}<0.24$ in the $p_{\rm init}=1$ models; 
$0.13<\Delta X/X_{\rm NEL}<0.17$ in the $p_{\rm init}=3/2$ models; 
$0.10<\Delta X/X_{\rm NEL}<0.14$ in the $p_{\rm init}=2$ models.
These differences do not show systematical dependence on $X_{\rm init}$: 
$\Delta X/X_{\rm NEL}$ decreases as $X_{\rm init}$ increases in the $p_{\rm init}=1$ models;
$\Delta X/X_{\rm NEL}$ increases as $X_{\rm init}$ increases in the $p_{\rm init}=3/2$ models;
$\Delta X/X_{\rm NEL}$ does not change systematically in the $p_{\rm init}=2$ models.

The $X/X_{\rm NEL}$ medians at $10^9$~yr are similar between the same $X_{\rm init}$ models.
Although photoevaporation strongly affects the $p_{\rm init}=1$ models, the smallest median at $10^9$~yr is 0.48 in the ($X_{\rm init}$, $p_{\rm init}$)=($0.1^2$,2) model.
This is because the $X/X_{\rm NEL}$ medians at $10^7$~yr right after the giant impact stage are the smallest in this model.
The envelope-losses via giant impact and photoevaporation are complementary: the envelope-losses via giant impact work efficiently on the massive planets and the envelope-losses via photoevaporation works favorably on the less massive planets.

\section{Comparison to observed planets} \label{sect:comparison}

Our results provide the orbital architecture of the formed planets, which include the size and size ratio distributions of adjacent planet pairs.
We compare these distributions to those of the observed planets.
We note that we do not consider the observation bias for the distributions of the observed planets.

\subsection{Period-Radius Distribution}

\begin{figure}
	\plotone{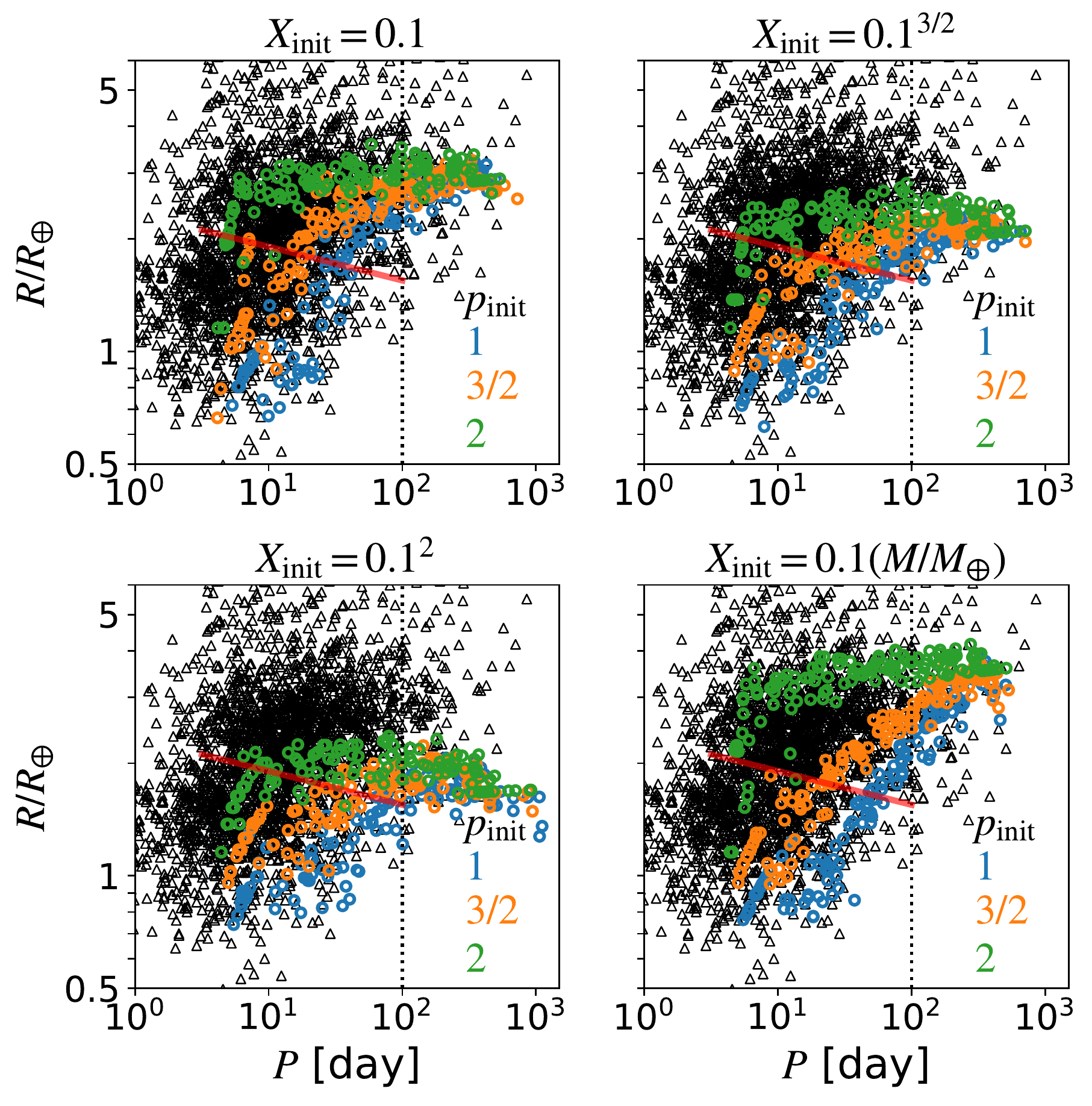}
	\caption{
		The $P$--$R$ distributions of simulated planets at $t=10^9$~yr (circles) and observed planets around F, G, and K stars (black triangles).
		For the simulated planets, different panels and different colors represent different $X_{\rm init}$ and $p_{\rm init}$ results, respectively.
		The observation data were extracted from the NASA exoplanet archive (https://exoplanetarchive.ipac.caltech.edu/) as of February 2021, and planets around F, G, and K stars are plotted.
		The red lines are the planetary size at the radius gap \citep[][]{Van_Eylen+2018}.
		}
		\label{fig:PR_psum_obs_PE_over2}
\end{figure}

\begin{figure}
	\plotone{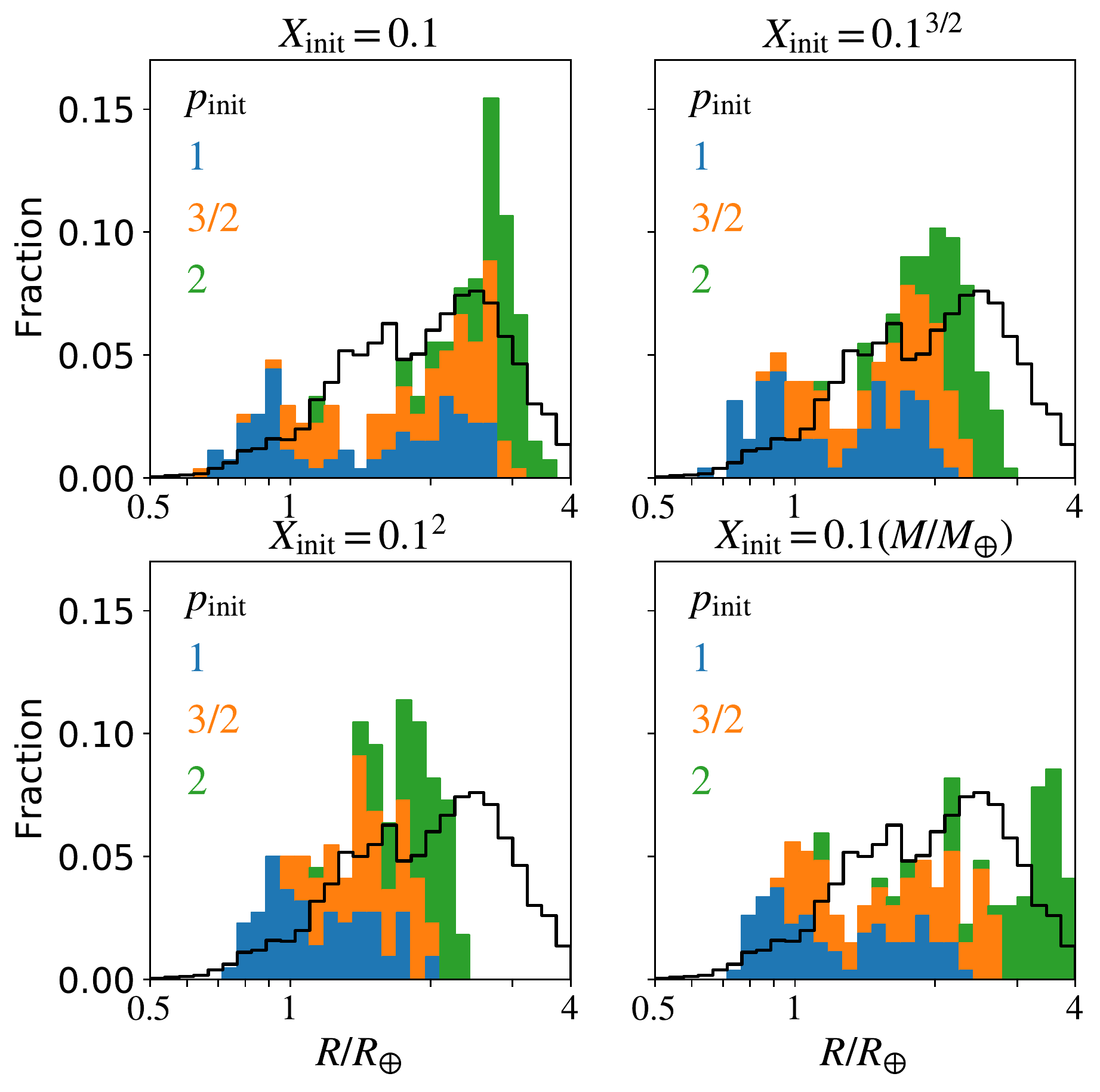}
	\caption{
		Histograms of the fractions of the simulated planets at $10^9$~yr (filled) and the observed planets around F, G, and K stars (open), using 30 logarithmic bins.
		These fractions are calculated from the planets whose periods are less than 100~days and sizes are between $0.5R_{\oplus}$ and $4R_{\oplus}$.
		We note that the fraction of the observed planets is not the same as the occurrence rate, i.e., the number of the observed planets per star \citep[e.g.,][]{Fulton&Petigura2018}.
		We do not consider the completeness correction since the low completeness of the small planets affects the fraction of the observed planets.
		}
		\label{fig:Rhist}
\end{figure}

First of all, we compare the simulated and observed planets on the period-radius plane (Figure \ref{fig:PR_psum_obs_PE_over2}).
We focus on planets with $\sim10$~day periods, of which the transit results are less prone to the survey incompleteness.
Most of the observed planets with $\sim10$~day periods are distributed between $R_{\oplus}$ and $4R_{\oplus}$.
This feature is reproduced when we consider the ensemble of the planets in the same $X_{\rm init}$ models.
If we consider the ensemble of the planets in the same $p_{\rm init}$ models, the size range of the observed planets with $\sim10$~day periods is not reproduced.
This is because given $p_{\rm init}$, the ensemble of planets populates in a narrow strip on the radius-period plane as illustrated in Figure \ref{fig:PE_PRX_XM} and explained in Section \ref{sect:result_PE}.
This indicates that the initial mass distribution is a possible explanation of the size distribution of the observed planets with $\sim 10$ day periods.
The observed $1R_{\oplus}$ sized planets are reproduced by the planets in the $p_{\rm init}=1$ and 3/2 models. 
The $4R_{\oplus}$ sized ones are reproduced by those in the ($X_{\rm init}$, $p_{\rm init}$)=(0.1, 2) and ($0.1(M_{\rm c}/M_{\oplus})$, 2) models.
The simulated planets with 10~day period orbits in these models are sparsely distributed.
The period-radius distribution of the observed planets would be reproduced if we perform more simulations for $p_{\rm init}$ to be continuously distributed from 1 -- 2 in the $X_{\rm init}=0.1$ or $0.1(M_{\rm c}/M_{\oplus})$ models.

\subsection{Radius Gap}

The radius gap is another important feature of the observed planets on the period-radius plane \citep[e.g.,][]{Fulton+2017, Fulton&Petigura2018}, 
as indicated by the red line in Figure \ref{fig:PR_psum_obs_PE_over2}, which generally separates two groups of planets -- the small inner planets and the large outer planets. 
Figure \ref{fig:Rhist} shows the size histograms of the observed planets (the open histogram) and simulated planets (the filled one), whose periods are less than 100~days and sizes are between $0.5R_{\oplus}$ and $4R_{\oplus}$.
Although the completeness correction is not applied, the observed planets show the gap at $R/R_{\oplus}\approx 1.8$, such as the data from the CKS survey showed \citep[][]{Fulton&Petigura2018}.
We note that the gap is deeper when the survey completeness is taken into account.
In our results, especially the $X_{\rm init}=0.1$ and $0.1^{3/2}$ models, the size distributions of the simulated planets are bimodal with deep gaps.
These two groups of planets are composed of the inner small-size planets, which have no or small envelopes, and the outer large-size planets, which have large envelopes.
These two groups are not clear in the $X_{\rm init}=0.1^2$ and $0.1(M_{\rm c}/M_{\oplus})$ models since the size difference between the inner small-size planets and the outer large-size planets is small in the $X_{\rm init}=0.1^2$ models and the sizes of the simulated planets are too diverse in the $0.1(M_{\rm c}/M_{\oplus})$ models.
The size distributions of the simulated planets in the $X_{\rm init}=0.1$ and $0.1^{3/2}$ models do not agree with that of the observed planets: the sizes of the two planet groups and the gap size are not consistent; the fractions of the outer large-size planets are higher than the observed fraction.
The planets around the gaps are contributed by the $p_{\rm init}=1$ and $3/2$ models (Figure \ref{fig:Rhist}).
The gaps in these models are located at $1.2R_{\oplus}$ -- $1.5R_{\oplus}$, which correspond to $2.1M_{\oplus}$ -- $5.1M_{\oplus}$ for bare planets.
These mass ranges are about equal to the masses of the massive planets with $\sim10$~day periods in the $p_{\rm init}=1$ and $3/2$ models (Figure \ref{fig:final_PMX_XM}).
The planetary size at the peak of the large planets on the histogram depends on $X_{\rm init}$: the peak is at $2.6R_{\oplus}$ in the $X_{\rm init}=0.1$ model and $2.0R_{\oplus}$ in $X_{\rm init}=0.1^{3/2}$ model.
For reference, this observed peak is at $2.4R_{\oplus}$ \citep[][]{Fulton+2017}, which is reproduced in \cite{Owen&Wu2017} when $X_{\rm init} \in [0.01,0.3]$ and the Rayleigh distribution of the planetary mass with a mode $3M_{\oplus}$.

The planetary size at the observed gap and fraction of planets that are smaller than the gap size would be possibly explained if we perform simulations when $1.5\leq p_{\rm init}\leq2$ in the $X_{\rm init}=0.1$ and $0.1^{3/2}$ models.
The gaps would move to a larger size since massive bare planets, which have larger masses than $5.1M_{\oplus}$, will form.
Planets in the $1.5\leq p_{\rm init}\leq2$ models also help to explain the fractions of the observed planets on the period-radius plane (Figure \ref{fig:PR_psum_obs_PE_over2}), where the simulated planets do not show the observed peaks in the population (see above and below the red line), and the histogram (Figure \ref{fig:Rhist}), where the simulated planets do not pile up around $1.3 R_{\oplus}$.

Additional envelope-loss mechanisms would help us explain the fraction of the observed planets that are larger than the gap size.
In this study, for example, 
we do not consider some envelope-loss mechanisms such as the Parker wind after giant impacts \citep[][]{Biersteker&Schlichting2019} and the core-powered mass-loss \citep[e.g.,][]{Gupta&Schlichting2019} to reproduce the radius gap.
These envelope-loss mechanisms make the fraction of the outer large envelope planets smaller, which helps to explain the fractions of the observed planets.

Although we consider the ensemble of the planets in the same $X_{\rm init}$ models, the distribution of the initial envelope fraction would be another possible explanation of the histograms of the observed planetary radii.
If we consider the ensemble of the planets in the ($X_{\rm init}$, $p_{\rm init}$) = (0.1, 2), ($0.1^{3/2}$, 2), ($0.1^{3/2}$, 3/2), ($0.1^{2}$, 3/2), and ($0.1^{2}$, 1), the size distribution becomes closer to the observed one.

\subsection{Planetary System Architecture}

\begin{figure*}
	\gridline{
		\fig{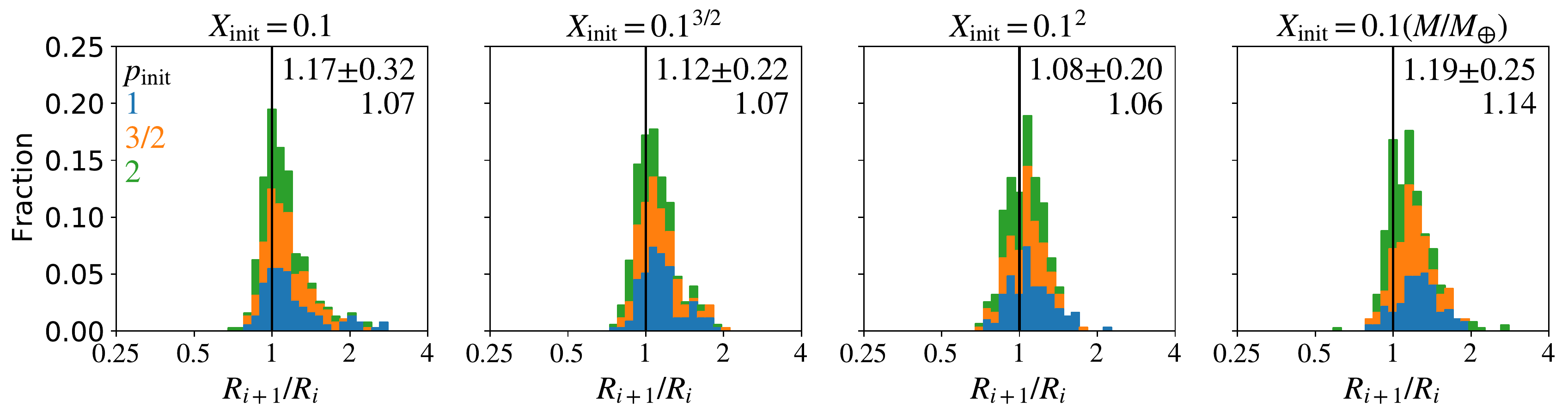}{1.\textwidth}
		{Top rows: The fractions of the size ratios between adjacent planet pairs using 40 logarithmic bins.}
		}
	\gridline{
		\fig{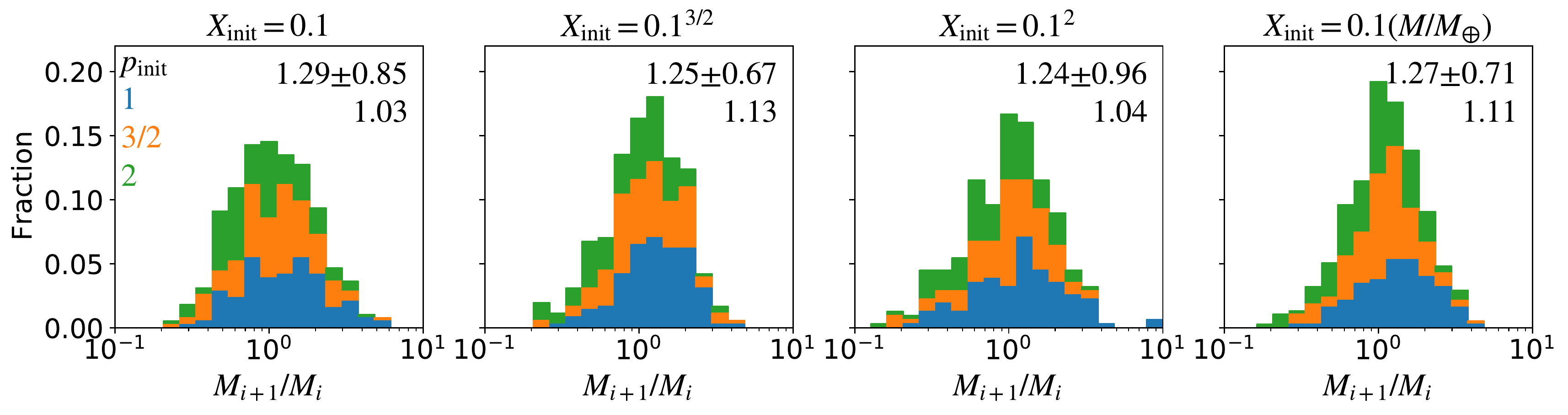}{1.\textwidth}
		{Middle rows: The fractions of the mass ratios between adjacent planet pairs using 20 logarithmic bins.}
		}
	\gridline{
		\fig{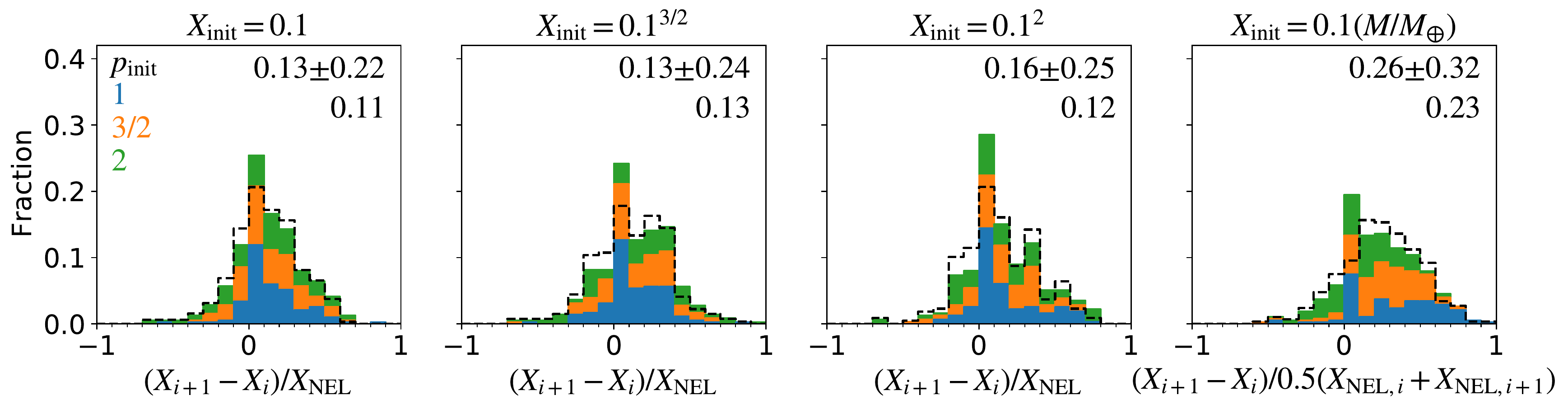}{1.\textwidth}
		{Bottom rows: 
			The fractions of the envelope fraction differences between adjacent planet pairs using 20 bins.
			The differences are normalized by $X_{\rm NEL}=X_{\rm init}$ in the constant $X_{\rm init}$ models and $0.5(X_{{\rm NEL},i}+X_{{\rm NEL},i+1})$ in the mass-dependent $X_{\rm init}$ models. 
			The dashed lines are the fraction excluding the $X_i=0$ pairs.
		}
		}
	\caption{
		Histograms of the fractions of the size ratios (top rows), mass ratios (middle rows), and envelope fraction differences (bottom rows) between adjacent planet pairs at $10^9$~yr.
		The upper right numbers in each panel are the average values, standard deviations, and median values, respectively.
		}
		\label{fig:Rrahist_psum_PE}
\end{figure*}

\begin{figure}
	\plotone{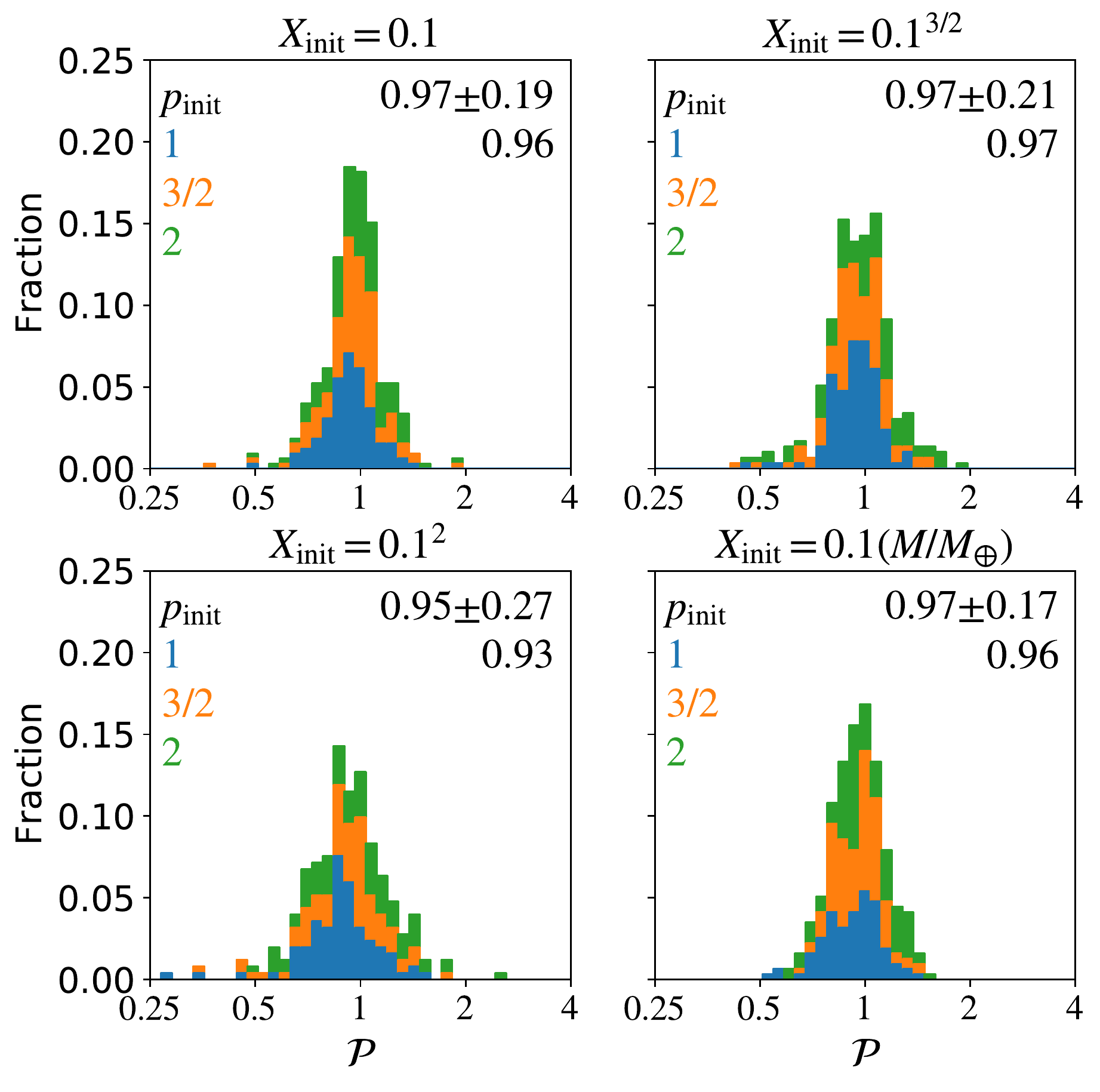}
	\caption{
		Histograms of the fractions of the ratio of the period ratios between adjacent planet pairs ($\mathcal{P}$) at $10^9$~yr, using 40 logarithmic bins.
		The upper right numbers in each panel are the average values, standard deviations, and median values, respectively.
		}
		\label{fig:Prasqhist_psum_PE}
\end{figure}

We focus on the sizes and periods of planets within each single multi-planet system.
The observed planets show that the average and standard deviation of the size ratios are $\langle R_{i+1}/R_i\rangle=1.29 \pm 0.63$ and their median is 1.14, and those of the ratio of orbital period ratios ($\mathcal{P}=(P_{i+2}/P_{i+1})/(P_{i+1}/P_i)$) are $\langle \mathcal{P} \rangle=1.03\pm0.27$ and their median is 1.00 \citep[][]{Weiss+2018}.
The size ratio distributions between the adjacent formed planet pairs at $10^9$~yr and the ratios of their period ratios are shown in Figures \ref{fig:Rrahist_psum_PE} and \ref{fig:Prasqhist_psum_PE}.
Both of them are peaked around 1, which indicates that the size and spacing of the adjacent planets are correlated and are consistent with the peas-in-a-pod pattern.
Comparing to the size ratio distribution of the observed planets, the standard deviations of the simulated planet pairs are small, which means that the size ratio distributions of the simulated planet pairs are more concentrated around 1.
The size ratios increase as $X_{\rm init}$ increases and $p_{\rm init}$ decreases.
The size ratio distributions are determined by the mass ratio and envelope fraction ratio distributions.
In Figure \ref{fig:Rrahist_psum_PE}, we also show the distributions of the mass ratios and the envelope fraction differences between the adjacent formed planet pairs.
The mass ratio distributions show the following features: the averaged mass ratios are about 1.3; most of the mass ratios are distributed between 0.5 and 2; the mass ratios tend to be high and their deviations are large in the small $p_{\rm init}$ models.
The masses of the adjacent planets are correlated since protoplanets grow by collisions with neighboring protoplanets (Section \ref{sect:N_par}).
Most mass ratios are in $0.5\lesssim M_{i+1}/M_i\lesssim 2$, which corresponds to $0.84\lesssim R_{i+1}/R_i\lesssim 1.2$ if we neglect the size of an envelope.
This lower value, 0.84, is almost consistent with the lower values of $\langle R_{i+1}/R_i\rangle$ of the simulated planets.
In contrast, the upper value estimated by the mass ratio, 1.2, is smaller than the upper values of $\langle R_{i+1}/R_i\rangle$ of the simulated planets since envelopes affect them.

The envelope fraction differences between adjacent planets normalized by the envelope fractions of the planet without envelope-loss are peaked at 0.
Their average values and dispersions are around 0.15 and 0.25 in the constant $X_{\rm init}$ models.
In these models, protoplanets initially have the same $X$, i.e., $X_{i+1}-X_i=0$.
Each planet becomes to have a different envelope fraction through giant impacts, $\langle X_{i+1}-X_i\rangle/X_{\rm NEL}=0.060\pm0.28$ in the $X_{\rm init}=0.1$ model, $0.074\pm0.31$ in the $X_{\rm init}=0.1^{3/2}$ model, $0.11\pm0.30$ in the $X_{\rm init}=0.1^2$ model.
Their averaged values are positive since outer planets tend to experience fewer collisions and have higher envelope fractions.
Photoevaporation affects the envelope difference as follows:
When the inner planets lose envelopes, the final envelope fraction differences become larger;
When both planets lose whole envelopes, $X_{i+1}-X_i=0$.
The final $(X_{i+1}-X_i)/X_{\rm NEL}$ is typically between 0 and 0.4, which contributes to $R_{i+1}/R_i$.
These average values become slightly small and the deviations are almost the same if we ignore the bare planet pairs.
In the mass-dependent $X_{\rm init}$ models, their average values and dispersions are larger than those in the constant $X_{\rm init}$ models.
Except for the $p_{\rm init}=2$ model, $(X_{i+1}-X_i)/0.5(X_{{\rm NEL},i}+X_{{\rm NEL},i+1})$ is initially non 0 value: 0.05~--~0.20 in the $p_{\rm init}=1$ model and 0.05~--~0.10 in $p_{\rm init}=3/2$ model.
After giant impacts, the average value is $0.25\pm0.33$, and there is no peak at 0.
The standard deviation is similar to those in constant $X_{\rm init}$ models after giant impacts.
The envelope fraction differences become peaked at 0 after photoevaporation. 

The average ratios of the period ratios between adjacent formed planets $\mathcal{P}$ are slightly smaller than 1, although that of the observed period ratios is 1.03.
The inner separations are slightly larger than the outer ones in our results.
The $\mathcal{P}$ values are small in small $p_{\rm init}$ models and $\langle\mathcal{P}\rangle=1.0$ in the $p_{\rm init}=2$ models.
In the small $p_{\rm init}$ models, inner protoplanets experience more collisions, which excite eccentricities (Section \ref{sect:N_par}), and as a result, the inner planets tend to have slightly large separations.
If we consider the continuous distribution of $p_{\rm init}$, the average $\mathcal{P}$ would be close to 1.

In summary, although the planet sizes can be reshaped by subsequent photoevaporation, the aforementioned size and period correlations start to develop during the giant impact phase. 
Our results suggest that the mass and envelope distributions of the initial protoplanets affect the final size and period ratio correlations.

\section{Conclusions} \label{sect:conclusion}

The Kepler transit survey with follow-up spectroscopic observations has revealed the orbital architecture of super-Earths and found interesting features of their period, size, and these ratio distributions.
To make the first attempt to explain all these distributions, we have investigated the size evolution of super-Earths by $N$-body simulations and subsequent photoevaporation simulations using the following simplified approach.
We start with the protoplanets, which are composed of cores and envelopes. 
We calculate their sizes by the summation of the sizes of cores and envelopes according to the analytical model by \cite{Owen&Wu2017}.
We consider the initial envelope fractions ($X_{\rm init}$) as a parameter prescribed by two models, which are the constant initial envelope fraction models ($X_{\rm init}=0.1, 0.1^{3/2}, 0.1^2$) and the mass-dependent envelope fraction model ($X_{\rm init}=0.1(M_{\rm c}/M_{\oplus})$).
Another parameter is the power-law index associated with the surface density of the initial protoplanets ($p_{\rm init}$).
Our main findings are summarized as follows:
\begin{enumerate}
	\item As giant impacts occur, eccentricities and inclinations are pumped up through scattering, leading to high-velocity collisions and sometimes causing efficient envelope erosions.
	\item Protoplanets at inner orbits experience more collisions, and as a result, inner planets sometimes become bare cores after giant impacts. The fraction of the bare core planets to the entire formed planets is less than 10\% and increases as $X_{\rm init}$ decreases. 
	\item 
	About 10\% of the entire formed planets keep the initial envelopes since they do not experience any collisions. 
	These initial-envelope planets are typically located at outer orbits. 
	They keep almost all envelope even after photoevaporation since their semimajor axes are $\gtrsim1$~au.
	\item While photoevaporation efficiently strips the gas envelopes from less massive planets at inner orbits, envelope-loss via giant impacts is effective for massive planets due to high collision velocities.
	\item Our results suggest that the period-radius distribution of the observed planets would be reproduced if we perform simulations with $1\leq p_{\rm init}\leq 2$ for either $X_{\rm init}=0.1$ or $0.1 (M_{\rm c}/M_{\oplus})$.
	\item The size distributions of the simulated planets are bimodal in the $X_{\rm init}=0.1$ and $0.1^{3/2}$ models.
	The small sized planets at inner orbits are produced by photoevaporation.
	However, the size distribution of the simulated planets is not consistent with that of the observed planets including the location of the radius gap. 
	The size distribution of the observed planets would be reproduced if we consider the continuous $p_{\rm init}$ between 1 and 2 and the envelope-loss mechanisms that we do not model in this study such as the Parker wind after giant impacts \citep[][]{Biersteker&Schlichting2019} and the core-powered mass-loss \citep[e.g.,][]{Gupta&Schlichting2019}.
	\item 
	The simulated planets show the peas-in-a-pod pattern, i.e., their size ratios and the ratios of the orbital period ratios between adjacent planets are around 1. 
	These distributions originate from the giant impact evolution between adjacent protoplanets, which have similar sizes, masses, and envelope fractions.
\end{enumerate}

We comment on our assumptions and future studies in the following. 
In this study, we have considered the effects of giant impacts and photoevaporation, separately.
While this two-step method is adopted for modeling simplicity, the envelope-loss timescale of less massive protoplanets at inner orbits via photoevaporation is comparable to the timescale of giant impacts.
More realistically, these effects work simultaneously.
We adopt a simple energy-limited escape approach for photoevaporation. 
However, some of the protoplanets have large sizes in their evolution, and these protoplanets would lose envelopes via the radiation-recombination-limited escape \citep[e.g.,][]{Murray-Clay+2009}.

We do not consider the thermal expansion of the envelopes after impacts.
If we consider the envelope expansion, we expect the following two effects: 
firstly, the envelopes are stripped \citep[][]{Biersteker&Schlichting2019} and secondly, the planets whose sizes are expanded tend to cause low-velocity collisions and keep slightly high envelope fractions (Section \ref{sect:N_std}, Figure \ref{fig:X_Xinit_cum_GIPE}).
These effects are opposite and the thermal evolution of the envelope after the impact would be a key to consider which is dominant.

We constrain the mass and envelope mass fractions of the initial protoplanets.
It is expected that these distributions are related \citep[e.g.,][]{Ikoma&Hori2012}.
Their realistic relationship would help us to constrain these initial distributions of the observed planets.
Besides, our simulations do not consider hit-and-run collisions.
If we consider hit-and-run collisions, it is expected that the number of collisions increase and the final planets have less envelopes.
These effects will be investigated in future works.

\acknowledgments

We thank the referee for helpful comments.
Numerical simulations and analyses were carried out on PC cluster at Center for Computational Astrophysics, National Astronomical Observatory of Japan and  at the Academia Sinica Institute for Astronomy and Astrophysics.
This research was supported by MOST in Taiwan through the grant MOST 109-2112-M-001-052.
E. K. is supported by JSPS KAKENHI Grant Number 18H05438.

\bibliography{bibtex_ym}{}
\bibliographystyle{aasjournal}

\end{document}